\documentclass[twocolumn,twocolappendix]{aastex63}
\usepackage{lineno}

\accepted{03/31/2021}
\shorttitle{THAI workshop report}
\shortauthors{Fauchez et al.}

\begin{document}

\title{TRAPPIST Habitable Atmosphere Intercomparison (THAI) \\ workshop report}

\thispagestyle{empty}

\author[0000-0002-5967-9631]{Thomas J. Fauchez}
\affiliation{NASA Goddard Space Flight Center
8800 Greenbelt Road
Greenbelt, MD 20771, USA}
\affiliation{Goddard Earth Sciences Technology and Research (GESTAR), Universities Space Research Association (USRA), Columbia, MD 7178, USA}
\affiliation{NASA GSFC Sellers Exoplanet Environments Collaboration}

\author[0000-0003-2260-9856]{Martin Turbet}
\affiliation{Observatoire  Astronomique  de  l’Universit\'e  de  Gen\`eve,  Universit\'e  de  Gen\`eve,  Chemin  des Maillettes 51, 1290 Versoix, Switzerland.}

\author[0000-0001-8832-5288]{Denis E. Sergeev}
\affiliation{Department of Mathematics,
College of Engineering, Mathematics, and Physical Sciences, University of Exeter
Exeter, EX4 4QF, UK}

\author[0000-0001-6707-4563]{Nathan J. Mayne}
\affiliation{Department of Astrophysics,
College of Engineering, Mathematics, and Physical Sciences, University of Exeter,
Exeter, EX4 4QL, UK}

\author[0000-0002-6776-6268]{Aymeric Spiga}
\affiliation{Laboratoire de M\'et\'eorologie Dynamique (LMD/IPSL), Centre National de la Recherche Scientifique,
Sorbonne Universit\'e, École Normale Sup\'erieure, \'Ecole Polytechnique, Paris, France}

\author[0000-0002-6673-2007]{Linda Sohl}
\affiliation{NASA Goddard Institute for Space Studies, New York, NY 10025, USA}
\affiliation{Center for Climate Systems Research, Columbia University, New York, NY, USA}

\author{Prabal Saxena}
\affiliation{NASA Goddard Space Flight Center
8800 Greenbelt Road
Greenbelt, MD 20771, USA}
\affiliation{University of Maryland
College Park, MD 20742, USA}
\affiliation{NASA GSFC Sellers Exoplanet Environments Collaboration}

\author[0000-0001-9423-8121]{Russell Deitrick}
\affiliation{Center for Space and Habitability, University of Bern, Gesellschaftsstrasse 6, CH-3012, Bern, Switzerland}

\author[0000-0002-2253-5802]{Gabriella Gilli}
\affiliation{Instituto de Astrofísica e Ciências do Espaço (IA), Tapada da Ajuda, Edificio Leste-2 piso, 1349-018, Lisbon, Portugal}

\author[0000-0003-0354-9325]{Shawn D. Domagal-Goldman}
\affiliation{NASA Goddard Space Flight Center
8800 Greenbelt Road
Greenbelt, MD 20771, USA}
\affiliation{NASA GSFC Sellers Exoplanet Environments Collaboration}
\affiliation{NASA NExSS Virtual Planetary Laboratory, Seattle, WA, 98195, USA}

\author[0000-0002-3262-4366]{Fran\c cois Forget}
\affiliation{Laboratoire de M\'et\'eorologie Dynamique (LMD/IPSL), Centre National de la Recherche Scientifique,
Sorbonne Universit\'e, École Normale Sup\'erieure, \'Ecole Polytechnique, Paris, France}

\author[0000-0003-3047-615X]{Richard Consentino}
\affiliation{NASA Goddard Space Flight Center
8800 Greenbelt Road
Greenbelt, MD 20771, USA}
\affiliation{University of Maryland
College Park, MD 20742, USA}

\author[0000-0001-6487-5445]{Rory Barnes}
\affiliation{Astronomy Department, University of Washington, Box 951580, Seattle, WA, 98195, USA}
\affiliation{NASA NExSS Virtual Planetary Laboratory, Seattle, WA, 98195, USA}

\author[0000-0003-4346-2611]{Jacob Haqq-Misra}
\affiliation{NASA NExSS Virtual Planetary Laboratory, Seattle, WA, 98195, USA}
\affiliation{Blue Marble Space Institute of Science, Seattle, WA, USA}

\author[0000-0003-3728-0475]{Michael J. Way}
\affiliation{NASA Goddard Institute for Space Studies, New York, NY 10025, USA}
\affiliation{NASA GSFC Sellers Exoplanet Environments Collaboration}
\affiliation{Theoretical Astrophysics, Department of Physics and Astronomy, Uppsala University, Uppsala, Sweden}

\author[0000-0002-7188-1648]{Eric T. Wolf}
\affiliation{Laboratory for Atmospheric and Space Physics, University of Colorado Boulder, Boulder, CO, USA}
\affiliation{NASA NExSS Virtual Planetary Laboratory, Seattle, WA, 98195, USA}
\affiliation{NASA GSFC Sellers Exoplanet Environments Collaboration}

\author[0000-0002-3249-6739]{Stephanie Olson}
\affiliation{Department of Earth, Atmospheric, and Planetary Science, Purdue University, West Lafayette, IN 47907, USA}

\author[0000-0003-2273-8324]{Jaime S. Crouse}
\affiliation{NASA Goddard Space Flight Center
8800 Greenbelt Road
Greenbelt, MD 20771, USA}
\affiliation{Center for Research and Exploration in Space Science and Technology II (CRESST II)/Southeastern Universities Research Association (SURA), 1201 New York Avenue, N.W., Washington, DC 20005}
\affiliation{NASA GSFC Sellers Exoplanet Environments Collaboration}
\affiliation{NASA NExSS Virtual Planetary Laboratory, Seattle, WA, 98195, USA}

\author[0000-0003-0475-8479]{Estelle Janin}
\affiliation{University College London, Mathematical and Physical Sciences faculty, Natural Sciences department}

\author[0000-0001-5657-4503]{Emeline Bolmont}
\affiliation{Observatoire  Astronomique  de  l’Universit\'e  de  Gen\`eve,  Universit\'e  de  Gen\`eve,  Chemin  des Maillettes 51, 1290 Versoix, Switzerland.}

\author[0000-0002-3555-480X]{J\'er\'emy Leconte}
\affiliation{Laboratoire d'astrophysique de Bordeaux, Univ. Bordeaux, CNRS, B18N, allée Groffroy Saint-Hilaire, Pessac, F-33615, France}

\author[0000-0003-4711-3099]{Guillaume Chaverot}
\affiliation{Observatoire  Astronomique  de  l’Universit\'e  de  Gen\`eve,  Universit\'e  de  Gen\`eve,  Chemin  des Maillettes 51, 1290 Versoix, Switzerland.}

\author[0000-0002-4842-1432]{Yassin Jaziri}
\affiliation{Laboratoire d'astrophysique de Bordeaux, Univ. Bordeaux, CNRS, B18N, allée Groffroy Saint-Hilaire, Pessac, F-33615, France}

\author[0000-0001-5328-819X]{Kostantinos Tsigaridis}
\affiliation{NASA Goddard Institute for Space Studies, New York, NY 10025, USA}
\affiliation{Center for Climate Systems Research, Columbia University, New York, NY, USA}

\author[0000-0001-6031-2485]{Jun Yang}
\affiliation{Dept. of Atmospheric and Oceanic Sciences, School of Physics, Peking University, Beijing, 100871, People's Republic of China}

\author[0000-0001-9771-7953]{Daria Pidhorodetska}
\affiliation{Department of Earth and Planetary Sciences, University of California, Riverside, CA, USA}
\affiliation{NASA Goddard Space Flight Center 
8800 Greenbelt Road 
Greenbelt, MD 20771, USA}

\author[0000-0002-5893-2471]{Ravi K. Kopparapu}
\affiliation{NASA Goddard Space Flight Center
8800 Greenbelt Road 
Greenbelt, MD 20771, USA}
\affiliation{NASA GSFC Sellers Exoplanet Environments Collaboration}
\affiliation{NASA NExSS Virtual Planetary Laboratory, Seattle, WA, 98195, USA}

\author[0000-0003-1995-1351]{Howard Chen}
\affiliation{Department of Earth and Planetary Sciences, Northwestern University, Evanston, IL 60202, USA}
\affiliation{Center for Interdisciplinary Exploration \& Research in Astrophysics (CIERA), Evanston, IL 60202, USA}

\author[0000-0002-1485-4475]{Ian A. Boutle}
\affiliation{Department of Astrophysics,
College of Engineering, Mathematics, and Physical Sciences, University of Exeter,
Exeter, EX4 4QL, UK}
\affiliation{Met Office, FitzRoy Road, Exeter, EX1 3PB, UK}

\author[0000-0002-3143-9716]{Maxence Lef\`evre}
\affiliation{Atmospheric, Oceanic and Planetary Physics, University of Oxford, Oxford, UK}

\author[0000-0003-0977-6545]{Benjamin Charnay}
\affiliation{LESIA, Observatoire de Paris, Université PSL, CNRS, Sorbonne Université, Université de Paris, 5 Place Jules Janssen, 92195
Meudon, France.}

\author{Andy Burnett}
\affiliation{Knowinnovation, 7903 Seminole Blvd, Suite 2303
Seminole
FL, 33772}

\author{John Cabra}
\affiliation{Knowinnovation, 7903 Seminole Blvd, Suite 2303
Seminole
FL, 33772}

\author{Najja Bouldin}
\affiliation{Knowinnovation, 7903 Seminole Blvd, Suite 2303
Seminole
FL, 33772}

\correspondingauthor{Thomas J. Fauchez}
\email{thomas.j.fauchez@nasa.gov}


\begin{abstract}
The era of atmospheric characterization of terrestrial exoplanets is just around the corner. Modeling prior to observations is crucial in order to predict the observational challenges and to prepare for the data interpretation. This paper presents the report of the TRAPPIST Habitable Atmosphere Intercomparison (THAI) workshop (14--16 September 2020).
A review of the climate models and parameterizations of the atmospheric processes on terrestrial exoplanets, model advancements and limitations, as well as direction for future model development was discussed. We hope that this report will be used as a roadmap for future numerical simulations of exoplanet atmospheres and maintaining strong connections to the astronomical community.
\end{abstract}

\keywords{Exoplanet atmospheres}

\section{Executive Summary} \label{sec:summary}
\subsection{Objectives} \label{subsec:obj}
The primary purpose of the workshop was to bring together a wide variety of participants in the exoplanet atmospheres community and beyond (Solar System and Earth Sciences) to discuss 3D general circulation models (sometimes also known as Global Climate Models, GCMs) in the context of exoplanet climates and atmospheric characterization. Specifically, the THAI project and workshop focused on modeling of TRAPPIST-1e, as it represents perhaps the best candidate for observation and atmospheric characterization of a terrestrial sized exoplanet in the habitable zone.  The THAI project was used as a vector for comparisons and discussions between the various GCMs that are currently commonly used for modeling terrestrial extrasolar planets. Particular attention was given to key parameterizations such as surface properties, moist convection, water clouds, radiative transfer, and non-LTE processes. Finally we also discussed how 1D models, such as energy balance models (EBMs) or single-column radiative-convective models, complement 3D models for exoplanet studies.

\subsection{Organization} \label{subsec:org}
Due to the COVID-19 pandemic the THAI workshop was held virtually, instead of in-person as was originally planned. The Scientifc Organizing Committe (SOC) consisted of Thomas J. Fauchez, Shawn D. Domagal-Goldman, Ravi Kumar Kopparapu, Linda Sohl, Martin Turbet, Michael J. Way and Eric T. Wolf. The THAI SOC worked with Knowinnovation (\url{https://knowinnovation.com/}), company led by Andy Burnett and assisted by Najja Bouldin and John Cabra, to build the conference website and to organize the live discussions. Each talk (twenty-six of about 12 to 15~min in length) was pre-recorded by the speakers and uploaded to the conference website at least a week before the live part of the workshop (September 14\textsuperscript{th} to 16\textsuperscript{th}). The talks are also permanently available on the  \href{https://www.youtube.com/channel/UCb0gqdGHntaPKxEuvc88Irg}{\textit{NExSS Youtube channel}}. The workshop attendees were therefore able to watch the presentations in advance and write questions to the speakers. Live sessions were limited to three hours per day (9~am to 12~pm EDT) divided in three parts: 1) questions and answer (Q$\&$A) session about pre-recorded talks, 2) coffee break in a 2D virtual reality space, 3) breakout discussions. Having an important part of the workshop offline helped to mitigate the impact of the time zone differences and travel issues, allowing more people, especially from underrepresented groups, to attend.

\subsection{Main Outcomes} \label{subsec:out}
This workshop's main scientific result is the intercomparison of four mature 3D GCMs used for simulating terrestrial climates; this will be presented in three separate papers as a part of a special issue of the Planetary Science Journal. 
During the workshop, the inter-model differences in the convection and cloud parameterizations have been highlighted as key culprits for disagreements between simulated climates. This necessitates future model development in this area of climate modeling, particularly given the importance of clouds and hazes in the observation and characterization of exoplanetary atmospheres. Furthermore, the dominance of one surface type or another --- e.g., ice, land, or ocean --- alters the planetary albedo which can significantly influence climate and habitability (Sec.~\ref{subsec:surf}).
Latitudinal EBM simulations either underestimate or overestimate  the strong day-night side contrast for synchronously rotating planets, although a longitudinal EBM can provide better representation of the temperature contrast between hemispheres (Sec.~\ref{subsec:syn}). A two column approach (day and night sides) shows promising results to capture the globally averaged surface temperature, and some degree of the hemispheric asymmetries. 
In the various discussions during the workshop, certain aspects of model intercomparison and potential areas of improvement were found to be shared with similar questions of modeling atmospheres of solar system planets, including Earth.
Concerns were raised over the carbon footprint of GCM simulations mostly due to the electricity demand of supercomputers. Performing runs responsibly and optimizing the GCM to reduce the computational time have been identified as the best mitigation strategy and this report also recommends that future studies that use GCMs evaluate the amount of CO$_2$ emissions related to the modeling activities and disclose it in the paper. However, those considerations should not prevent researchers to perform the numerical experiments required by their science investigations.
Finally, we need to advance aspects of diversity, inclusivity, belonging, and justice in the field. This will require multiple efforts at both the individual level, and at the group and community levels. The long-term positives from such efforts will improve both the community we do our research within and the products from that community.

This workshop report is structured as follow. In section \ref{sec:overview} we introduce the THAI project and how Earth and Solar System intercomparisons can help us to build meaningful ones for the exoplanet community. In section \ref{sec:predict_obs} we discuss how GCMs are crucial to predict and interpret exoplanet observations. We then review in section \ref{sec:GCMparam} GCM parameterizations for exoplanets, their limits and the developments needed. In section \ref{sec:survey} we show the results of a survey filled by the workshop participants concerning the future of exoplanet GCMs. We follow in section  \ref{sec:div} by discussing diversity, equity and inclusion in the community. Finally conclusions and perspectives are given in section \ref{sec:end}.

\section{General Discussions about  Intercomparisons} \label{sec:overview}
In this section we present the THAI project and we discuss how current Earth and Solar System intercomparisons can help us to build successful ones for exoplanets.

\subsection{GCM intercomparisons for Earth and beyond} \label{subsec:inter}

No object in space is more well studied and has as significant a dataset regarding its extant and past state as the Earth.  Indeed, much of the understanding of planets in our solar system and exoplanets has been informed by principles gleaned from the study of different processes on the Earth.  This is especially true when considering the impact of geophysics on the modeling of exoplanets using GCMs. 

The development of GCMs for Earth Science studies has enabled their use for other planets. But, there are also key ways in which exoplanet GCM simulations can in turn improve our understanding of the Earth system and its evolution. At a fundamental level, exoplanet GCM simulations inherently act as a means of stress testing and phase space exploration that can then be applied to Earth-tuned models.  Due to the wide range of conditions that may exist on exoplanets, GCM simulations are typically run with lower complexity than leading Earth-tuned models, but explore conditions over a wider phase space that may lie at the limits of the model's physical validity.  The results from these models can expose minor bugs or clarify the effects of varying specific parameters that can potentially feed back to Earth-tuned models and the geophysics assumptions that underpin them.  

The lesser expense of these simpler exoplanet GCM simulations and their tendency to explore this wider phase space also has a more direct influence on the understanding of the effect of certain features of geophysics on Earth.  Simulations that explore the general effects of variations in bulk geophysical parameters or of external parameters are now often run for a range of exoplanet parameters or for a broad suite of terrestrial planets \citep{Wolf2017, Way2018}.  These simulations and their findings can then provide a library of outcomes and lessons that can be applied to specific geophysical conditions on simulations of the extant Earth or for paleo-Earth climate simulations. 

In this subsection, we review briefly some of the highlights of previous Earth, Solar System and exoplanet intercomparisons.

\subsubsection{The TRAPPIST Habitable Atmosphere Intercomparison (THAI)} \label{sec:thai}

Preliminary results of the THAI intercomparison \citep{Fauchez2020THAI} have been published. Four atmospheric compositions have been simulated by the four GCMs (see Sec.~\ref{subsec:GCMs}). The simulated atmospheres include two dry (no surface liquid water) benchmark cases ``Ben1" and ``Ben2", presented by Martin Turbet (\url{https://www.youtube.com/watch?v=B8a2-G8NmmA}), with the atmospheric compositions of 1 bar of N$_2$, 400 ppm of CO$_2$ and a purely 1 bar of CO$_2$, respectively, and two moist habitable cases ``Hab1" and ``Hab2", with a global ocean and the same respective atmospheric compositions, presented by Thomas J. Fauchez (\url{https://www.youtube.com/watch?v=kYLbp_BrJFs}). The overall outcome of the Ben1 and Ben2 cases is a good agreement between the four GCMs in terms of surface and atmospheric fields. However, we note some differences in the circulation regime, which manifest a sensitivity of TRAPPIST-1e simulations to GCM setup due to a combination of planetary parameters, noted e.g., in \citet{Sergeev2020}. Synthetic transmission spectra have been produced by the Planetary Spectrum Generator \citep[PSG,][]{Villanueva2018} and they are in good agreement between the models as long as the top of the atmosphere is extended (assuming an isothermal atmosphere and fixed gas mixing ratio) up to about 100 km (10$^{-7}$ and 10$^{-10}$ bar for Ben1 and Ben2, respectively). Without this extension, the strongest absorption features of CO$_2$ are truncated. 
Concerning the Hab1 and Hab2 cases,  clouds are the largest source of discrepancies between the models, as expected, due to differences in convection (Sec.~\ref{subsec:conv}), bulk condensation, cloud microphysics, boundary layer, and other parameterizations, and their coupling with atmospheric dynamics and radiation. The altitude and thickness of the cloud deck at the terminators impact the simulated transmission spectra, leading to about 20\% differences between the models on the number of transits required to detect those atmospheres with James Webb Space Telescope (JWST) at a 5$\sigma$ confidence level.
More details on the Ben1 \& Ben2 simulations, Hab1 \& Hab2 simulations and on the impact on observable transmission spectra and thermal phase curves will be presented in three follow-up papers.
We also welcome other modelling groups to join THAI at any time.
Interest has been shown from the THOR \citep{Deitrick2020} and Isca \citep{Vallis2018} GCM groups that we hope to host and compare results soon, while the Exeter Exoplanet Theory Group (EETG) will contribute results from the UM's replacement, LFRic \citep{Adams2019}, in the future.

Once JWST data for TRAPPIST-1e are available, GCM output will be compared to observational data. We expect such comparison will lead to a new set of simulations and to further validation of model performance against terrestrial exoplanet data.
It is important to maintain and improve the level of collaboration between the exoplanet GCM community (including THAI) and the observational community. The ``TRAPPIST-1 JWST Community Initiative'' \citep{gillon2020trappist1} is particularly  relevant as it aims to develop a coordinated framework to study TRAPPIST-1 planets with JWST from the both the observational and theoretical/modeling levels.

\subsubsection{Model Intercomparisons Across Stellar Spectral Types \citep{yang_differences_2016,Yang2019}}

\cite{yang_differences_2016} compared the differences in 1D radiative transfer calculations among two line-by-line codes (SMART and LBLRTM), a moderate resolution code (SBART) and four low-resolution codes that are used in GCMs (CAM3, CAM4\_Wolf, LMD-G, and AM2).  Note CAM4\_Wolf would eventually become ExoCAM (see Appendix \ref{subsubsec:ExoCAM}).  The atmospheric composition was set to 1 bar of N$_2$, 376 ppmv CO$_2$, and variable H$_2$O. They showed that there are small differences between the models when the surface temperature is lower than about 300 K. At higher temperature, such as 320--360 K, the differences between the models could be tens of watts per square meter. The source of the differences are mainly from water vapor radiative transfer calculations in both shortwave and longwave. The differences are larger for shortwave than longwave and also for an M-dwarf spectrum than a solar spectrum. These results suggest that radiative transfer codes should be verified first (such as the absorption and continuum behavior of water vapor) before being used in an exoplanet GCM especially when targeting planets with hot climates or to estimate the inner edge of the habitable zone.  Notably, an important lesson learned from this study is that the adequate performance of shortwave radiative transfer for warm moist atmospheres is contingent upon sufficiently resolving the near-IR H$_2$O spectral absorption bands and windows, particularly when considering irradiation from M-dwarf stars.

\cite{Yang2019} compared the differences of 3D GCM simulation results on a rapidly rotating aqua-planet receiving a G-star spectral energy distribution and on a tidally locked aqua-planet receiving an M-star spectral energy distribution. Several GCMs were considered, various versions of CAM, the LMD-G and the AM2 GCM \citep{GFDL2004} They found  relatively small difference ($<$8~K) in global-mean surface temperature predicted by various GCM for cloudy planets orbiting a G star but rather large differences (20-30~K) for cloudy planets orbiting M stars. These discrepancies are due to differences in the atmospheric dynamic, clouds and radiative transfer. Clouds is the largest difference between the models. The interactions between radiative transfer (such as shortwave absorption by water vapor) and atmospheric circulation can influence the atmospheric relative humidity and therefore affect the surface temperature.

\subsubsection{GCM intercomparisons for Solar System planets}
%
\paragraph{Mars.} Several GCM intercomparison efforts were arranged in the Mars atmosphere modelling community as soon as enough teams could contribute to such a project. They were organized in advance of two workshops that helped structure the community just like the 2020 THAI workshop did, although no official reports were published in the literature. The first meeting was the ``Mars GCM intercomparison workshop'' organized at Oxford University, United Kingdom, July 22-24, 1996. It was later followed by the first ``Mars atmosphere modelling and observations'' which took place in Granada (Spain) on January 13-15, 2003. In both cases instructions were sent to the different teams to prepare comparable simulations in advance, and volunteers analyzed the simulations by comparing the zonal mean fields and the predicted planetary waves. Radiative transfer models were also compared for specific test cases \citep[as announced by][]{Harri2003}. These efforts showed that, most of the time, significant differences could always be attributed to different settings in the models, and that it was difficult to reach profound scientific conclusions from such organized intercomparisons. Mars GCM teams later focused on comparing their models with the numerous observational datasets that became available in the 2000s. Nevertheless specific intercomparison studies have continued being conducted to study phenomena or altitude ranges for which not much observational data are available \cite[e.g.,][]{Gonzalez2010,Wilson2014} or to compare model predictions at a mission landing site \cite[e.g.,][]{Kass:03,Newman2020}.

\paragraph{Venus and Titan.} Similarly, intercomparison campaigns have been organized in the Titan and Venus GCM communities which reached a sufficient stage of development a few years after the Martian case. In practice, because on Venus and Titan the problem of superrotation is so striking and challenging for the GCMs, the most interesting comparisons actually focused on the behaviour of the various dynamical cores and their ability to simulate superrotation and conserve angular momentum \citep[][see also section~\ref{sec:gcm-known-limi}]{Lebonnois2012,Lebonnois2013}. These studies revealed that various dynamical core which would give very similar
results in Earth or Mars conditions, can predict very different circulation
patterns in Venus-like conditions.  
Recently, a detailed Titan climate model intercomparison has been organized, with the motivation of preparing the planned Dragonfly Titan lander mission \cite[]{Lora2019}, like intercomparison studies done for Mars to prepare for a mission landing. On the basis of the acceptable agreement between the different models, the authors conclude that the ``low-latitude environment on Titan at this season is now fairly well constrained'', which is reassuring when preparing an ambitious mission like Dragonfly.

\paragraph{Snowball Earth.} A noticeable GCM intercomparison has been presented in \cite{Abbot:2012} concerning the impact of clouds in the snowball Earth deglaciation. Six different GCMs were onboard the intercomparison. They found that clouds could warm a snowball Earth enough to reduce the amount of CO$_2$ required for deglaciation. But because the amount of clouds varies from one model to another, the amount of CO$_2$ required differs by one order of magnitude depending on the model. This intercomparison highlights clouds as an important source of discrepancies between the GCMs.

\subsection{Ideas for Advancing Exoplanet Model Intercomparisons}
\subsubsection{The Coupled Model Intercomparison Project (CMIP) as a Guide for ``ExoMIPs"}\label{subsubsec:CMIP}

In planning future community-based exoplanet model intercomparisons (``ExoMIPs") like the THAI project, it is useful to consider the 25-year history of the future Earth-focused Coupled Model Intercomparison Project (CMIP). This discussion has been presented at the THAI workshop by Linda Sohl.

CMIP is perhaps best known now for its contributions to the periodic assessments issued by the Intergovernmental Panel on Climate Change (see, e.g., IPCC report 2013, \url{https://www.ipcc.ch/report/ar5/wg1/}), but it began in 1995 as an independent project of the World Climate Research Program (WCRP). Over the years, CMIP has grown, from an effort to use simple global coupled ocean-atmosphere GCM experiments, with  interactive sea ice and land models, to separate natural climate variability from human-induced climate change \citep{Covey2003}, to the exploration of a variety of sophisticated climate change scenarios, using GCMs with ever-more complex capabilities that include chemistry and higher-resolution dynamical interactions \citep{Eyring2016}, as well as specialized ancillary investigations (e.g., the various projects under the \href{https://pmip.lsce.ipsl.fr/}{\textit {Paleoclimate Modelling Intercomparison Project, or PMIP}}). The history of the CMIP experience highlights four key considerations that would benefit any future exoMIP effort:

\textit{Context/Rationale: Establish how the MIP will advance the state of knowledge and/or the state of the art.} CMIP’s overall experiment designs are developed as an outgrowth of the \href{https://www.wcrp-climate.org/grand-challenges/grand-challenges-overview}{\textit {WCRP Grand Challenges}}, which are updated periodically via community input. These Grand Challenges, which encompass observational, theoretical, and modeling-based research, are meant to: 

 \begin{itemize}
     \item Identify the key research questions needing to be addressed in order to move the field forward in a substantive way, as well as the barriers to progress (what do we need to learn next, and what stands in the way?); 
     \item Define effective and measurable performance metrics (how will we know we have been successful in achieving our goals?);  
     \item Provide storylines that engage a broad interested audience, from the media and general public to scientists from other disciplines (how can we attract future talent and improving interdisciplinary connections?).
\end{itemize}

Planetary science and astrobiology do not have an internationally defined set of grand challenges as such. However, documents such as the \href{https://astrobiology.nasa.gov/research/astrobiology-at-nasa/astrobiology-strategy/}{\textit {NASA Astrobiology Strategy}} and \href{https://www.liebertpub.com/doi/full/10.1089/ast.2015.1441}{\textit {AstRoMap European Astrobiology Roadmap}} outline research topics of interest for advancing the field, and not surprisingly there is overlap that can help narrow the context and rationale of an exoplanet MIP. An ExoMIP should strive to make connections with as many of these topics as is plausible: linking the MIP rationale to broad themes of community-wide interest is to our advantage in connecting to our fellow researchers whose focus is on theory, field work or observations.

\textit{Experiment Design: Encourage broad participation by planning core experiments with low entry barrier for most groups, with specialized subprojects as needed.} One of the benefits of MIPs, from a model development standpoint, is that comparisons across multiple models can illustrate which model design/parameterization approaches provide the most robust results. Thus the more models we can encourage to participate in a given MIP, the better for the community. The participation of any one modeling group in a MIP is going to be limited by three factors: the technical requirements of the MIP (how intensive is the set-up process for the experiments?); the available resources (how much computing time is available?); and expertise on hand (are there enough people with the necessary model knowledge and time to run all the experiments?). 

CMIP has taken the approach of defining a limited set of required core experiments that are easy to implement \citep[e.g.,][]{Eyring2016}. This lowers the entry barrier to groups interested in joining, and improves the chances of useful outcomes beneficial to the community. More complex scenarios, specialized topics, and extended parameter space are addressed through \href{https://www.wcrp-climate.org/modelling-wgcm-mip-catalogue/modelling-wgcm-cmip6-endorsed-mips}{\textit{related MIPs}}, which attract groups that have the additional resources, interest and expertise available. For an ExoMIP, it is conceivable that a similar design could focus on global-scale core experiments with simple forcing changes (stellar insolation, relatively thin atmospheres/oceans), and some of the related MIPS could engage with 1D/EBM models on details of atmospheric composition/radiative forcing.

\textit{MIP Logistics: Plan realistic schedules for experiment completion and group analyses/manuscript preparation.} Keeping model groups focused and reaching project milestones in a timely fashion are important for the overall success of a MIP. CMIP and the related MIPs typically establish these schedules via community planning workshops, where ideas for experiments and additional ``rules of engagement" regarding MIP participation are also defined in advance. These MIP protocols should then be published as close to the official start of a MIP project as possible, so that additional groups not involved in the planning workshops can also make a timely decision to participate. 

\textit{Data Sharing: When/how to release experiment results for broadest impact?} On this topic, CMIP and some of the related MIPs - mainly the PMIP projects - handle data sharing differently. Because the groups contributing to CMIP are frequently not entirely the same as researchers conducting multi-model evaluations of the experiment results, data are released immediately to the community. In contrast, the specialized PMIP project experiments are often run and evaluated by the MIP participants themselves. In the latter case, a data embargo is often declared until the first group papers are published, as part of the agreed-upon project schedule. An ExoMIP might want to consider a similar data embargo, as part of a community agreement not to publish each other’s work prematurely.

Data sharing logistics are a more complicated issue. No ExoMIP will have the vast resources currently available to CMIP for data sharing, so at present any data sharing is likely to happen on an ad hoc basis. However, it is possible to develop community standards for what and how data should be shared. While raw model output and some accompanying post-processing scripts might provide maximum flexibility to fellow modelers, model results that have already been post-processed into file formats such as netCDF for map views and plain text for line plots allows the broadest possible audience - from fellow scientists who are not modelers, to educators and students - to work with and learn from the output with the help of free apps.

\subsubsection{Planning Workshop Themes for Rocky Exoplanet Model Intercomparisons} \label{subsec:challenges}

The beginning of the first network of exoplanet model intercomparison is a great time to address organizational challenges, and it should be a top priority for the community. 

In order to advance inter-model comparisons for exoplanets study, a first-order requirement is a collaboration workshop for the community to discuss key issues, especially regarding data to be shared (and how to share it), which would be very important for establishing standards and best practices for intercomparisons going forward. \textit{A formal intercomparison workshop could be organized for roughly one year after the THAI workshop (fall 2021), which would allow time for planning the workshop and getting funding support to encourage participation. }

We can expect that an intercomparison workshop would produce documentation about best practices for model intercomparisons, and most importantly, produce a community consensus on how to share data so that a common repository (to be identified) would host a comprehensive set of common diagnostic outputs that are most important for addressing the science questions asked (e.g., which diagnostic attributes contribute most to the synthetic spectral signatures of interest for observational programmes?), and yet do not overwhelm potential users or the repository itself with ancillary diagnostics/large data files.

 To compare GCMs, even to one another, we must necessarily address a range of cases (by bracketing single-point cases with increasingly complex physics included sequentially, etc.). This requires a clearly defined question (e.g., ``is this climate sensitive predominantly to clouds or surface water reservoir effects") with the goal of producing concrete test cases (that is, with an exhaustive list of clearly stated parameters) and testable predictions, so that it is possible to differentiate possible states. Having a clear, well-defined question, in conjunction with a concise list of simulations and deliverables, is necessary to ensure that requirements for participation can be met within realistically allocatable work-efforts for such projects. Transparency with respect to set parameters also ensures reproducibility, and provides diagnostic access for future tests and observations. Future observational modes further constrain important test deliverables (wavelength coverage and spectral resolution if spectroscopic, or dynamic/geologic/phenomenological diagnostics; spatial scales; arrival times; duration of observation or mission lifetime, etc.). If the models are too far away from one another, then how do each of the GCMs motivate those disparate results, and does this lead to additional testable predictions? 

Given the abundance of ``hidden" parameterizations in models, it can be difficult to assess whether a given model in a given part of parameter space matches observations for the ``right" reason (e.g., the same physical driver as the primary control in both the model and the planetary environment), versus a confluence of other effects. Further intercomparison work can elucidate some of these factors, but it is unlikely that they will all become explicit dependencies, even with additional documentation. Note that potential new observables outside current capabilities and ones where additional precision would refine parameter ranges also helps to steer future mission development, which then feeds back into how close the models are to ground-truth. Continued intercomparisons require continued funding, and given the dependence of adequate documentation, this suggests the need for clear funding lines for development of testing frameworks (either for a single model or as part of a new or ongoing collaboration), validation, documentation, etc.  GCM model development work is costly, and national laboratories like NCAR or GISS typically hire software engineers to support the scientists.  The exoclimate community is young and the scientists generally lack such support, hampering progress.  Lastly, reducing the model output to a single metric (e.g., generic climate state or spectral signature produced) is an additional constraint (meta-sensitivity?), suggesting that disagreeing models may ``agree" in some diagnostic sense. This would help to identify additional directions to explore, such as broadening the parameter space identified initially, or through secondary observables (diurnal/seasonal variability, etc.). 

Finally, we would also recommend that potential intercomparison contributors think beyond the goal of ``what does this planet simulation look like from the observation perspective?" 3D GCMs in particular produce a wide variety of diagnostic outputs that are interesting and relevant to understanding the potential habitability of a particular world configuration, as well as model performance, but these do not necessarily produce directly observable results. This is especially true knowing that some observables that we have currently identified may be in fact unobservable and that new ones will eventually be found later. Current modeling should therefore not be only constrained by the current set of possible observables.

\subsubsection{Discussing and building the Climates Using Interactive Suites of Intercomparisons Nested for Exoplanet Studies (CUISINES)} \label{subsec:CUISINES}
In the upcoming era of JWST, it becomes essential to focus community effort on benchmarking/comparing/validating the performance of exoplanet climate models, both with respect to other models and to observations (when available). As noted in Section \ref{subsubsec:CMIP}, model intercomparisons have been widely used for decades by the Earth science community in this way, as a very valuable means to improve model reliability, mitigate model dependencies,  track down bugs, and provide benchmarks for new models. While individual intercomparison projects should have their own clearly defined protocols, the exoplanet community would benefit also from a metaframework -- essentially, a framework for designing model intercomparison projects. This metaframework is what we propose with CUISINES. This framework would be open from 0D to 3D models as well as radiative transfer models, and not limited only to rocky exoplanets. One of the first steps in establishing this metaframework will be to create a CUISINES committee and then to prepare a workshop on best practices for intercomparisons. 

At the end of the THAI workshop, two future intercomparisons were already discussed: one between EBMs for ice belts, and one between GCMs for cloud-free mini Neptunes (see Section \ref{subsubsec:beyond} below). In the era of JWST, mini-Neptunes and hot Jupiters in particular will require focused modeling efforts from the community. Other ideas of intercomparisons under the CUISINES metaframework are welcome.

\subsubsection{Moving Beyond Rocky Planets: Envisioning a Mini-Neptune Model Intercomparison}\label{subsubsec:beyond}
A future THAI-equivalent GCM intercomparison project for mini-Neptunes has been proposed during a breakout discussion. Mini-Neptunes are the most abundant category of exoplanets that have been discovered so far and thanks to their larger size they will be more easily characterized through transmission spectroscopy with JWST. For now, there is a wide range of approaches considering mini-Neptunes modelling so there is a significant need for an intercomparison to see what the differences are when everyone makes the same assumptions (more important for now than thinking about the more complicated physics we need to implement).
For instance, aerosols are very challenging to be included in mini-Neptunes \citep{Charnay:2015a,Charnay:2015b,Charnay:2020}, therefore it has been suggested to consider clear-sky simulations as a start. Cloudiness has been shown to decrease with decreasing equilibrium temperature \citep{Crossfield2017} which motivates the case for relatively cold temperatures \citep[K2-18b][and colder]{Benneke2019,Tsiaras2019} where less photochemical haze is expected. Gliese-436b could be a very good candidate with respect to the amount of data potentially available and the quality of constraints on the planetary parameters \citep[and references therein]{Gillon2007,Demory2007,Lanotte2014,Ehrenreich2015,Bourrier2016,dosSantos2019}. Atmospheric compositions should be limited to common gases expected for these planets such as hydrogen, helium, water, methane and carbon dioxide and surface pressures could range from few millibars to tens of bars. 
However, in order to take into account deep atmosphere effects on the upper atmosphere, that have been shown to be important for Titan \citep{Lebonnois2012} it may require inclusion of pressures up to 10 to 100 bar which can subsequently increase the computational time \citep{Wang:2020}. 
 In a next step the atmospheric compositions can be refined to include more processes to match with observations that JWST would have provided. Also other models such as EBM and 1D radiative-convective models could be engaged too.

\section{GCM Simulations to Predict and Interpret Exoplanet Atmospheric Characterization} \label{sec:predict_obs}


The strength of the exoplanet modeling community is in its close connection between scientists from numerous disparate fields, including astronomers, climate scientists, planetary scientists, and geophysicists.  Knowledge from all of these fields should be leveraged when considering exoplanetary atmospheres, and inputs from each field can be incorporated into GCM simulations of exoplanetary climates.  Due to the computational expense of GCMs, before conducting large sets of simulations, we tightly constrain the goals of our simulations sets.  Given the observational limitations that will persist through the coming decade, from an astronomical perspective it would be most beneficial to prioritise categories of planets that are going to be definitively observable in the near future.
We should compile a list of potential targets in the order of importance --- this would help to distribute our computing resources more efficiently.
Model development effort should then mainly be directed towards the most observable types of planet. For example, can we observe a Mars-like planet tidally locked to an M-dwarf star? Or are exo-Venuses our most realistic target? Astrobiological implications is an obvious motivation for simulating exoplanet atmospheres. However the characterization of Earth-sized habitable worlds is far more challenging (and perhaps untenable) compared with characterization of sub-Neptunes and larger worlds. One strategy for constraining habitable zones and habitability may lie first, paradoxically, in constraining uninhabitable regions in planet-phase spaces, thus allowing us to eliminate planets from the list of potentially habitable worlds.
In other words, we have to be able to readily distinguish extreme atmospheric conditions from Earth-like atmospheres.

The simplest answer to what planets and parameter space will be a target of near term observations are planets we can actually observe.  Characterization of exoplanets is still challenging for most planets smaller than large Jupiter-sized planets and the ability to characterize smaller planets is one of the highly anticipated open areas that JWST will explore.  While the highest quality data from JWST with respect to characterization will be for hot Jupiters, there is still significant interest in characterizing rocky planets, if feasible, with JWST and ELTs. The recognition that characterization of potentially rocky planets may be the next frontier that is explored in exoplanet science is what has motivated much of the recent study of theoretical models of rocky planets. These theoretical studies of the different potential environments of rocky exoplanets and their potential observational signatures are key to ensuring that sufficient understanding exists for interpretation of observations of rocky exoplanets when they are obtainable. In the intermediate range between Jupiter sized planets and potentially rocky Earth sized planets are the class of planets that may be the next truly characterizable planets in the near term, so called ``mini-Neptunes" and ``super-Earths".  These terms are meant to describe the mass and radius regime of these worlds as there is considerable uncertainty regarding their interior, surface and atmospheric properties.  However, it is precisely this uncertainty and the potential to extricate key parameters from observations of these worlds that makes them potentially valuable probes of planetary formation, evolution and habitability. Aside from the abundance of potential targets, Super Earths and small Neptunes also inhabit a potentially critical region of planetary parameter space. These planets likely bracket the point at which runaway accretion of a primary gaseous atmosphere occurs in the core accretion model \citep{Pollack1996}. Therefore, they bridge the structures of giant planets with thick hydrogen/helium dominated atmospheres, to terrestrial planets with much thinner ``secondary" atmospheres \citep{Lopez2014}, as well as being in the size range where irradiative evaporation becomes significant \citep{Owen2012}.  

The array of observational techniques that will be used to extricate the properties of these worlds are both those that have been heavily used in the past and that are in the nascent stages of being leveraged. For the latter, emission spectroscopy may be a critical means by which to probe into deeper portions of the atmosphere, despite the presence of expected clouds and hazes. Ground-based observations will also be key, particularly high-resolution spectroscopy -- possibly coupled with direct imaging -- that may be diagnostic of composition and other atmospheric features using some of the large planned near term ground-based telescopes \citep{Snellen:2015,Lovis:2017}.  More familiar data products such as light curves will continue to be critical as their multi-wavelength morphology will be key to informing and validating climate models.  

This connection and feedback between observations and theoretical work will be key to near term interpretation of exoplanet observations in the context of atmospheric and surface characterization. Synergies between retrievals and GCMs will enable the connection of the existing physical and chemical models to observations in way that may be able to elucidate parameters that will be informative regarding planetary formation and evolution. To facilitate this, there exists the need for closer connection between chemistry models and GCM models \citep[e.g.,][]{Chen2019,Drummond_2020}.  In addition to that, there is a need for generalized condensation schemes (for a broad range of planets from hot rocky planets to cold gaseous planets and for a variety of atmospheric compositions).  Connections to and integration of other key modeling, such as modeling of atmospheric escape and the evolution of planets given different formation pathways, will also need to supported. Observations will drive much of this theoretical work, and both cooler ``sub-Neptunes"/``super-Earths" and hot rocky planets that may be characterizable by JWST are key.  

While connecting simulations to observations is key, another essential component of exoplanet characterization is the need to be confident in our models through validation practices such as inter-comparisons like THAI before applying them to understand terrestrial exoplanet observations.  The validation practices will be especially important because modeling of some of the most favorable near term observational targets will require the addition of novel functionality in a number of areas.  The following are just some areas that will likely require model development in order to appropriately model exoplanets that are likely to be near term observational targets:

\begin{itemize}
    \item Modelling ``sub-Neptunes"/``super-Earths" with extended atmospheres will require deep atmosphere equations, as the primitive equations break down due to thickness of atmosphere relative to planetary radius \citep{Mayne_2019}.
    \item The range of atmospheric compositions that will have to be considered will also expand for planets that are not $H_{2}$-rich or Earth-like or that do not have any representation in our Solar System \citep{Woitke2021}.
    \item Updated chemistry schemes that capture non-equilibrium or photochemistry effects that are likely relevant for these classes of exoplanet will be required.
    \item There is a need for an improved understanding of interior mixing to get proper boundary conditions for GCMs.
    \item There will be a need to be able to run simulations at lower pressures in order to properly treat photochemical hazes and other upper atmosphere processes that may affect observables.
    \item Robust parameterizations for convection that can deal with non-dilute condensibles will be required.
\end{itemize}    

Deserving of their own section (i.e. section \ref{subsec:aerosol})  clouds and hazes are the elephant in the room for improved understanding of characterizable exoplanets.  Observations of planets in and outside our solar system indicate that understanding of clouds and aerosols are required to even have a first order understanding of a planets' extant state and consequently its evolution. While modeling of clouds and hazes are inherently a complexity-rich endeavor that require trade offs due to limits of understanding or computational complexity, there are some key needs that will be required in order to accurately characterize exoplanets in the near term.  Understanding the inherent variability of a planet and determining the level at which weather and cloud variability changes the continuum level of observations is important for both interpretation and planning of observations.  The coupling of clouds and photochemical hazes to dynamics is also important to determine impact on transmission, emission and reflection spectra and phase curves (relevant for effectively all classes of planets, as demonstrated by solar system objects). An additional near term focus will be the inclusion of coupling detailed cloud microphysics codes (e.g., Helling code, CARMA, EddySed) to GCMs as has been done for hot Jupiters \citep{2018A&A...615A..97L}. The addition of models of increasing complexity, such as these, will likely require a model hierarchy going between simple models and the coupled cloud microphysical models. 

For planets in more extreme temperature regimes, such as hot rocky planets amenable to JWST characterization, there will be a need for significant model updates including updated radiative transfer (RT) schemes and non-dilute condensables development. These worlds will also require 1D models to take into account surface chemistry due to potential magma oceans. Finally, additional factors such as gravity waves will require appropriate parameterizations, since their impact in the upper atmosphere is significant for planets of the solar system \citep[e.g.,][]{Lott2013, Hoshino2013, Gilli2020} and potentially for hot Jupiters (for example, see \citealt{watkins_2010}).

Addressing these questions, we must design our numerical experiments efficiently.
To do large sets of simulations, we can adopt statistical approaches to cover as much of the parameter space with as few simulations as possible \citep[e.g., Latin Hypercube, see][]{Sexton2019}.
For example, we can use a decision tree of specific well-known biosignature parameter sweeps for each stage of Earth's history.
Parameter space can be covered efficiently also by relying on synergy between EBMs and 3D GCMs following an asynchronous coupling.
Namely, a ``rough" climate state can be spun up by a resource-cheap EBM and then explored in more detail with a resource-expensive GCM, followed by another spin-up period done with the EBM, and so on.

We believe the same methodology should be considered for future coupled atmosphere-ocean simulations of exoplanets, in which the ocean part requires longer timescales than the atmosphere.
We also discussed the untapped computational potential of graphical processing units (GPUs), which are currently underused by the GCM community \citep[e.g., THOR, see][]{Mendonca2016b,Deitrick2020}.
Future GCMs should ideally be developed agnostic of the machine architecture \citep[see][for example]{Adams2019}.
Using GPUs and similar hardware optimized for heavy computations would help to run large parameter sweeps in less time.

To summarize, there are a lot of necessary planet types to simulate and a lot of new couplings between atmosphere and other processes (e.g., atmosphere-ocean) to explore.
Future model intercomparisons should focus on relatively more observable atmospheres, keeping close connection with observational data.
We have to be careful in selecting modeling targets: on one hand it is more interesting to run simulations of exotic (relative to Earth) atmospheres; on the other hand these atmospheres are notoriously difficult to simulate with Earth-tuned codes, leaving very few GCMs being able to join the intercomparison.
Looking in a more distant future (the following decade perhaps?), a new generation of GCMs should be developed to be able to simulate such extreme cases as non-dilute, fully collapsible, or non-ideal-gas atmospheres.

\section{GCM Parameterizations, Limits and Development Needed}\label{sec:GCMparam}
\subsection{Sensitivity to numerical settings and initial conditions}
\subsubsection{Horizontal numerical diffusion}
Most GCMs require a numerical diffusion or filter which is applied in addition to the existing terms of the Euler or primitive equations \citep[][Chapter 13]{Lauritzen2011}. This mechanism typically serves two practical purposes that are intimately related: providing numerical stability, and achieving a kinetic energy spectrum that is consistent with our understanding of turbulent cascades.

In the case of numerical diffusion, diffusivity is generally a tune-able parameter. For Earth GCMs, the diffusivity can be tuned to achieve a spectral slope that matches empirically measured values \citep{Nastrom1985,Lauritzen2011}. For exoplanets, there is little hope of measuring the kinetic energy spectrum, but we can use the expectation of a $-3$ power law \citep{Charney1971} or $-5/3$ power law \citep{Pope2000,Caba:20} as guidance.

More specifically, the turbulent cascade causes energy to build up at the grid scale, well above the level at which molecular viscosity would act to convert this energy to heat \citep[see][Figure 13.7]{Lauritzen2011}. A numerical diffusion term is thus usually included, and selected to have a form that preferentially diffuses the fields at the smallest scales in the model by using, for example, iterated Laplacian operators \citep[see e.g.,][appendix A.2]{Spiga2020}. However, selecting the strength and form of this term is somewhat of an art, as it will depend on the resolution, time step size, solver, grid, and numerous other factors \citep{Lauritzen2011,Thrastarson2011}. Fortunately, most  exoplanet observables are relatively insensitive to the strength of numerical diffusion \citep{Heng2011,Deitrick2020}. Nonetheless, the exoplanet GCM community should bear in mind that other properties may be sensitive to ad hoc numerical settings. For instance, differences have been noted between GCM prediction of Titan superrotation possibly due to numerical diffusion \citep{Newman2011}.

\subsubsection{Sponge layers} \label{subsubsec:sponge}
One numerical issue deserves further attention: the need for so-called ``sponge layers'', that is, enhanced diffusion or drag near the model top and/or bottom. This need arises because GCMs typically use reflecting boundary conditions, allowing waves (usually gravity waves) to be reflected back into the model domain. These reflected waves are unphysical and can amplify and trigger numerical instabilities \citep{Lauritzen2011}. Thus an additional drag mechanism is often used to eliminate these reflections. Various types of sponge layer exist, for example, Rayleigh friction, which directly damps wind speeds toward zero or another value such as the zonal mean \citep[see, for example,][]{Mayne_2014,Mendonca2018}. This type of sponge is easy to implement but is non-conservative
\citep[for instance, terrestrial studies by][]{Shaw:07} show that sponge layers adversely impact the angular momentum balance, thus the simulated circulations). 
It is instructive to note that some GCM simulations of solar-system gas giants do not employ a sponge layer so as to avoid altering the angular momentum balance of the atmosphere \citep{Schn:09,Liu:10jets,Spiga2020}.
Another commonly used sponge layer, which is conservative in finite-volume formulations, reduces the order of the numerical diffusion, from (for example) fourth order to second order \citep{Lauritzen2011}. In either case, it is not always clear how to tune the strength and size of the sponge layer, which needs to damp waves without strongly affecting the general circulation or inducing additional reflections. 
While sponge layers have been carefully calibrated for Earth simulations, the effects of sponge layers and reflected waves arguably deserves more attention in exoplanet atmospheres.
Indeed, the exact settings required for these various damping mechanisms are currently unknown, as there is little constraint from observations, therefore, although in cases physically motivated (e.g., capturing dissipation from sub--grid eddies, or emulating the propagation of waves into space) their main use is for numerical stability \citep[see][for an example of the dissipation and maximum wind speed]{heng_2011}. 

\subsubsection{Initial conditions}
There has been some debate in the literature on the sensitivity of hot Jupiter simulations to initial conditions \citep{Thrastarson2010,Liu2013}. Other recent work has hinted at the possibility that zonal wind speeds on these planets may be sensitive to the initial temperature-pressure profile used in the deep atmosphere \citep{SainsburyMartinez2019}. More investigation should be done on initial conditions in exoplanet GCMs, although it is a challenge to explore many possibilities with such computationally expensive models. 

\subsubsection{Conservation properties}

The atmospheres of exoplanets present new territories in which physical processes may be unfamiliar and poorly constrained, compared to Earth and other solar system bodies. Unlike bodies in our solar system, for exoplanets we have little, if any, spatial information on the atmospheric structure. Thus in modeling these atmospheres we must utilize any and all available criteria to ensure physical realism.

The dynamical cores of GCMs are formulated using conservation laws (e.g., the Euler equations). As such, the global conservation of properties such as mass, energy, and angular momentum provides a diagnostic of the model's performance and accuracy. \cite{Thuburn2008} provides some guidance on which properties may be conserved and the desirable degree of conservation. We reiterate a few of those concepts here but more details on the dynamical cores will be given in section \ref{subsec:limdyn}. 

Firstly, as pointed out in \cite{Thuburn2008}, while the continuous forms of the equations of fluid dynamics can be formulated to conserve all physical properties, the discrete forms of the equations do not. Choices must be made regarding which properties to conserve (to numerical precision). One example of this is the thermodynamic equation (i.e., the first law of thermodynamics), which can be written in terms of different variables, such as potential temperature, pressure, internal or total energy, etc. Many GCMs use potential temperature, which is convenient for modeling convective processes. This lead to a conservation law for entropy, whereas sometimes a conservation law for energy (total or internal) may be preferred \citep{Satoh2002}. A similar choice must be made in the momentum equations, as these may be written in terms of linear momenta, angular momenta, vorticity, etc.

Conservation of mass is particularly important, as noted by \cite{Thuburn2008}, since it affects all other conservation laws, and should be robust in the absence of significant sources and sinks (e.g., escape to space or volatile freezeout). In other words, errors in mass conservation will lead to a cascade of errors elsewhere. 

As we develop GCMs further, and explore novel territory, conservation also provides a critical way to identify coding errors. In his talk, Russell Deitrick briefly discussed using mass and angular momentum conservation to identify bugs in the THOR GCM. Finite volume models (such as THOR), should conserve properties naturally to roughly machine precision because the equations are discretized in flux form---fluxes flow across boundaries such that control volumes on either side experience the exact same flux. Model discretized in other ways (e.g., spectral models) may ensure conservation by use of ``fixers'' \citep[see][Chapter 13]{Lauritzen2011}.

\subsubsection{Grid choice}

Traditionally, GCMs have used a latitude-longitude spherical grid, which is easy to construct and has operators that are well-known and intuitive. It does, however, suffer from singularities and resolution clustering due to the convergence of meridional lines at the poles \citep{Staniforth2012}. Many models have solved this issue using a combination of semi-implicit time integration and numerical filters, for example the UM. 

Another solution is to use an alternative horizontal grid structure. A comprehensive review of grid types, and their advantages and disadvantages, is provided in \cite{Staniforth2012}. The most commonly used of these seem to be the cubed-sphere, used in versions of the FMS, for example \cite{Lin2004}, and the icosahedral grid, used in NICAM \citep{Tomita2004}, THOR \citep{Mendonca2016b}, and DYNAMICO \citep{Dubos2014}. These quasi-uniform grids succeed in making the resolution more uniform across the sphere, avoiding numerical complications at polar regions in the lat-lon grid. These grids also scale very well with a high number of processors, which is usually not the case for the lat-lon grid \citep{Staniforth2012}. However, these grids are a true challenge to work with---the divergence, gradient, and curl operators must be written using Gaussian integrals on the icosahedral grid, for example \citep{Tomita2004}. Further, they are not \emph{completely} uniform and thus still admit the possibility of grid imprinting, wherein errors build up to a level that makes the underlying grid visible to the eye \citep{Staniforth2012}. Also, the core utilization of isocahedrad grid is far bellow the one of lat-lon grid. Unfortunately, there is no known ``perfect'' grid, so an understanding of the shortcomings of a particular grid is essential, as is comparison between models utilizing different grids. 

A further word of caution is warranted here: some models, as in \cite{DobbsDixon2008}, have avoided numerical issues with polar regions by omitting them entirely. While some features of the resulting simulations may be qualitatively reasonable, it is likely that the polar regions are particularly crucial for the circulation of tidally-locked exoplanets.


\subsection{The Limits of Dynamical Cores for Exoplanets} \label{subsec:limdyn}

One of the goals of atmospheric modeling for exoplanets
is, obviously, to unveil and to disentangle
the physical processes
underlying the observed properties.
Another goal relates directly to the science
of modeling itself: using hydrodynamical solvers
(dynamical cores) and subgrid-scale parameterizations
in the extreme conditions of exoplanetary atmospheres
is interesting because it could illustrate
the limitations of dynamical cores in a two-fold perspective
\begin{enumerate}
    \item Exoplanets allow us to explore from a fresh perspective known limitations of atmospheric modeling encountered in Earth and solar system planet applications (section~\ref{sec:gcm-known-limi}).
    \item Exoplanets can be very exotic (when considered from a solar-system-centered point of view) therefore new limitations arise when applying atmospheric numerical models to these environments (section~\ref{sec:gcm-new-limit}). 
\end{enumerate}
%
Before delving into a description of those challenges and limitations, it is important to note that, despite these challenges, e.g., studies of hot Jupiters using GCMs have provided excellent insight, including an almost complete picture of the acceleration of the zonal flow \citep[e.g.,][]{showman_2011,hammond_2020,Debras_2019,Debras_2020}, departures from chemical equilibrium caused by 3D dynamical mixing \citep[e.g.,][]{Drummond_2020}, and potential trends and characteristics of clouds \citep[e.g.,][]{lee_2016,Lines_2018,parmentier_2020}.
Hot Jupiters have been targeted as the main exoplanets to which GCMs have been applied due mainly to them being the most observationally-constrained cases as far as exoplanets are concerned.

\subsubsection{Known limitations considered with a fresh perspective \label{sec:gcm-known-limi}}

\paragraph{Dissipation and accuracy} When GCMs are used to model more ``extreme" planets, such as hot Jupiters, changes in the physical conditions can lead to reductions in the stability and potential accuracy of the simulation results. As discussed GCMs rely on several forms of dissipation in the model \citep{Jabl:11} to control sub-grid ``noise" which can lead to model instability. These can take several forms, from a diffusion of the winds themselves, ``filtering" over polar regions in latitude-longitude grid GCMs and so--called ``sponge" layers (see section \ref{subsubsec:sponge}, paragraph on sponge layers).

\paragraph{Rayleigh drag and closing the angular momentum budget} 
The question of the conservation of axial angular momentum is paramount in atmospheric modeling, as is illustrated for instance in studies of the slow-rotating bodies and gas giants in the solar system \citep[their appendix]{Lebo:12super,Spiga2020}. Dynamical cores are not explicitly formulated to conserve axial angular momentum and this can cause spurious variations of modeled angular momentum that can range from negligible to major, as was evidenced e.g., in the case of exoplanet modeling by \cite{Poli:14}.
This question of angular-momentum conservation is all the more critical in planets without a solid surface: a simple Rayleigh drag on horizontal winds is used as a bottom boundary condition in GCM studies of Jupiter and Saturn to emulate the closing of angular momentum budget in simulated jets by putative magnetic drag at depth \citep{Liu:10jets,Youn:19partone,Spiga2020}.
In the simulations of hot Jupiters too, Rayleigh drag, used as a simple parameterisation of surface drag on horizontal winds, has been used to capture the impact of magnetic drag in the deep atmosphere of hot Jupiters \citep{perna_2010}. However, in a similar fashion as the above-mentioned dissipation, its main use is for stability as by dragging the deep atmosphere to immobility where the radiative timescale is long, or indeed infinite \citep{iro_2005}, one can remove dependency of the simulated results on the initial conditions \citep[see][ for various discussions on this issue]{Mayne_2014,Amundsen_2016,Tremblin_2017,Sainsbury_2019}. Early simulations with the {\sc MITGCM} demonstrated a loss of global axial angular momentum without this inner boundary drag \citep{polichtchouk_2015}, and similar effects have been found for both {\sc UM} and {\sc THOR}. However, often the angular momentum conservation may vary with the spatial and temporal resolution adopted for model simulations; this opens the possibility to obtain negligible changes in axial angular momentum with adjustment to the spatial and temporal resolution of the particular setup. The cause of the AM conservation issue in dynamical cores is still not clearly understood. 

\paragraph{Thermodynamics} In most of the existing GCMs for planetary atmospheres, molecular weight gradients, heat capacity, gravity etc, are not taken into account. Those quantities are therefore assumed constant through the whole atmosphere and through time. This assumption can have several effects: when the thickness of the atmosphere becomes large with respect to the radius (such as for Titan or mini-Neptunes), the mass of a given atmospheric cell should change through buoyancy, due to the change of gravity. A constant or variable value of the heat capacity at constant pressure $C_p$, would directly impact the stability profile of the atmosphere. Such effects play on atmospheric dynamics and therefore on the equilibrium between the different terms of the atmospheric equations, but are generally assumed to be second order effects (except when the variations are really strong).
Taking into account this variability into GCMs would require subsequent development and time, and this endeavour would benefit both solar system planets and exoplanets. Preliminary works on how to take into account variations of $C_p$ with the temperature have already been presented in \cite{Lebonnois2010} and \cite{Mendonca2016}. 

\subsubsection{New limitations, specific to exoplanets \label{sec:gcm-new-limit}}

\paragraph{Shocks}
For some simulations the flow speed can approach, or exceed, a Mach number of one, leading to some authors questioning whether shock capturing solutions to the continuity equation are required \citep{Li2010,fromang_2016}. However, as shown by \citet{fromang_2016}, shocks play a minimal role in atmospheres dominated by a large-scale superrotating jet. As is understood for the solar system's gas giants, flows of conductive material in the presence of a background magnetic field can lead to drag (as mentioned earlier) and heating through ``ohmic dissipation" \citep[see for example][]{ginzburg_2016}. In hot Jupiters, the outer layers can become ionised, leading to magnetic drag, and the deeper layers potentially experience significant enough ohmic heating to alter the planetary radius. To date, the only simulations consistently capturing these impacts have been those of \citet{rogers_2014,rogers_2014b} which revealed that ohmic heating is unlikely to be significant enough, for reasonable magnetic fields. Magnetic drag, aside from the work of \citet{rogers_2014,rogers_2014b} has otherwise been captured through parameterised drag schemes \citep{perna_2010}.

\paragraph{Sub-stellar objects} 
As more and more planets have been detected and other observable populations identified, either through discovery, or improvements in instrumentation, the use of GCMs for gas giant extra-solar objects has widened. Brown dwarfs, sub-stellar objects which form similarly to stars, share parameter space with gas giant planets in terms of bulk compositions (to some extent) and radii, leading to similarities in atmospheric circulations in the two kinds of objects \citep{Show:19}. Additionally, due to high levels (relative to Jovian planets) of interior convection these objects are self--luminous, and high--signal to noise data is available \citep{buenzli_2015}. GCMs have been applied to these objects exploring their flows and additionally the occurrence of clouds \citep[e.g.,][]{zhang_2014,tan_2020}. The same studies can be applied to young gas giant planets, with high interior convection (e.g., directly imaged planets). The main challenges are in handling the strong interior fluxes from the convection, and the extremely short rotation periods. Work is beginning on coupling models of the convective interior of gas giant planets to atmospheric models, to better capture the interaction between these two regimes. Irradiated brown dwarfs, with a partner star from either the main sequence or white dwarf \citep{casewell_2018} have also recently been studied using an adapted GCM \citep{lee_2020}. Through studying this collection of gas giant objects, from young self-luminous Jovian exoplanets, to older short and long-period Jovian planets, and isolated or heavily irradiated brown dwarfs, a complete continuum of atmospheric regimes could be unraveled.

\paragraph{Adaptations related to exotic thermodynamics and chemistry} Observations of hot Jupiters have also begun to demarcate this sub--class itself into further categories. In particular, ultra-hot Jupiters (with temperatures in excess of $\sim$2,500\,K) provide some real advantages, whilst presenting new challenges. Although the relatively high temperatures result in the assumption of chemical equilibrium holding over most of the observable portion of the atmosphere, effects of the high temperature and photon fluxes such as thermal and photo dissociation, and the resulting $H^{-}$ opacity must be included \citep{baxter_2020}. Most significantly, these objects span a temperature range in which hydrogen is present in both molecular and atomic forms \citep{Bell2018,Tan2019}. The variations this causes in the specific heat capacity are large enough to mean the standard assumption of a single value through the atmosphere, made within GCMs, may become problematic. Resolution of this issue requires a significant reworking of the dynamical cores developed under the assumption of constant heat capacity.

\paragraph{Specific challenges for Mini-Neptunes} The drive in instrumentation is pushing towards detection and characterisation of smaller radii and longer orbital period planets, ultimately in the search for potentially habitable planets (the focus of this workshop). However, the next set of observational facilities and instruments will likely provide access to the sub-class of planets discussed previously, termed mini-Neptunes or Super--Earths. The existing challenges listed in section~\ref{sec:gcm-known-limi} clearly apply to those objects. Nevertheless, for these planets, initial work has shown that the standard primitive equations of motion often employed within a GCM may not be valid \citep{Mayne_2019}, and/or elapsed simulation times must be significantly extended \citep{huize_2020}. Additionally, as temperatures cool, the role of photochemistry and condensation may become even more important, but for a range of species. 

As highlighted by \cite{leconte_condensation_2017}, a background gas lighter than the condensible gas (for example H$_2$O and H$_2$) can induce a mean molecular weight gradient which inhibits convection in the atmosphere. This process is stronger on giant planets such as Mini-Neptunes but could be observed on smaller rocky planets. Therefore, it may be interesting to explore this phenomenon with 3D simulations.

\subsection{Parameterization of Convection in Exoplanet GCMs, differences and limitations} \label{subsec:conv}

\subsubsection{General considerations}
Convection, and moist convection especially, is an important driver of heat redistribution in planetary atmospheres.
By forming clouds and depending on the boundary-layer processes, moist convection is also a key part of complex feedback mechanisms in the climate system \citep[e.g.,][]{Arakawa2004}.
To accurately resolve convective plumes, a numerical climate model has to have sufficiently high spatial resolution, making it extremely computationally expensive and thus unfeasible for long simulations or multi-planet studies.
Thus, all modern exoplanet GCMs rely on parameterizations to emulate the overall effect of subgrid-scale convective processes on large-scale atmospheric fields.
These parameterizations always include a number of quasi-empirical parameters, usually inherited from Earth climate models or validated against observations and convection-resolving simulations on Earth, raising the question of their applicability to extraterrestrial atmospheres.

For the Earth's atmosphere, an assumption of dilute condensible is usually a good approximation \citep{Pierrehumbert2010}.
This is not the case for other planetary atmospheres in the Solar System and beyond: the main condensible species can comprise a substantial portion of the atmosphere or have thermodynamic properties such that the convective mass-flux \citep{Ooyama1971} is sufficient to affect the large-scale dynamics, like on Pluto \citep{Bertrand:2018}. In such a scheme, the depth of the convection is constrained by the distance the air, rising in convective updrafts, penetrates above its level of neutral buoyancy and ceases to rise.
To simulate these effects correctly, the LMDG is equipped with a convection parameterization accounting for non-trace condensible species \citep{Pierrehumbert_Feng2016}. It is applicable to a wide range of atmospheric conditions and is described in \cite{Leconte:2013b}.

Whatever the convection parameterization is based on --- the adjustment to reference profiles, subgrid-scale mass-flux  or other principles --- it has to be validated against observations and convection-resolving simulations.
In the absence of in-situ observations for exoplanets, the next best option is to use high-resolution convection-resolving and cloud-resolving models (CRMs), which simulate convective processes explicitly.
Targeted limited-area CRM simulations can be used to benchmark and improve convection parameterizations. For instance, \cite{Abbot2014} have compared CRM simulations to GCM simulations in the case of a snowball Earth and found that they provide consistent results. This helped to confirm the hypothesis that clouds could provide a large warming on a snowball Earth and potentially lead to the deglaciation of the planet.
Two talks at the THAI workshop presented work paving the way in this direction.
Denis Sergeev demonstrated substantial differences between cloud profiles in a global coarse-resolution GCM experiment and a limited-area CRM experiment, conducted using the UK Met Office Unified Model (\ref{subsubsec:UM}) for THAI Hab1 \& Hab2 setups.
Differences in convective cloud cover appear to be model- and planet-dependent, because the opposite picture has been found by Maxence Lef\`evre, who compared LMD-G GCM (\ref{subsubsec:LMD-G}) to the (Weather Research \& Forecasting) WRF model \citep{Skam:08} in a CRM mode for a case of convection on Proxima Centauri b.
Further work will hopefully build on Sergeev's and Lef\`evre's CRM simulations to explore convective processes in other atmospheric and planetary regimes, informing atmospheric modellers of parameterization biases and caveats.

\subsubsection{Transitioning from mass-flux scheme and convective adjustment toward fully resolving the convection: challenges and potential science returns.} \label{subsec:mf2ad}

It was a general consensus among the participants that a flexible convection parameterization based on the mass-flux approach is the best option for coarse-resolution 3D GCMs, while the adjustment scheme is usually too crude to represent convection. 
A shift towards fully-resolved global convection simulations is not going to happen quickly, but limited-area CRMs should be used more actively in the exoplanet atmospheric modelling.
As outlined in Sec.~\ref{subsec:conv}, one of the main applications of CRMs is to re-tune existing convection parameterizations for different extraterrestrial atmospheres.
Such experiments are routinely performed for the Earth weather and climate prediction models \citep{Rio2010} and for Mars as well \citep{Col2013}, so the exoplanet GCM community should work closely with meteorologists and Earth model developers to benefit from their invaluable expertise.
In practical terms, an important and already feasible project is building an archive of convection-resolving simulations of H\textsubscript{2}, N\textsubscript{2}, CO\textsubscript{2}-dominated atmospheres, which then can serve as a standard benchmark suite for coarse-grid global models.
This project can then evolve into an exoplanet CRM model intercomparison project and find its rightful niche under the CUISINES umbrella (Section \ref{subsec:CUISINES}).
With a wide grid of CRM models at hand, certain aspects of convective processes can then become more tractable, such as whether the structure and dynamics of convective plumes and precipitation anvils in non-Earth planetary atmospheres might change. How might they be influenced by plume size, convective overshoot beyond the level of neutral buoyancy, entrainment and detrainment, timescales of convection change when the composition of the atmosphere or stellar/planetary parameters (such as stellar spectrum and planetary gravity) are changed.

In addition, Earth science expertise is valuable for the development of generalized convection schemes from scratch.
In this case, code developers should strive to make them flexible, modular and portable so that it will be relatively easy to swap one convection parameterization in the GCM for another.
The fact that most convection schemes usually operate column-wise without communicating with neighbouring columns in a 3D GCM, makes the issue of portability easier to tackle.

\subsection{Planetary surface parameterizations for exoplanets and their impact on the climate}\label{subsec:surf}

\subsubsection{Land} \label{subsubsec:land}

There are many ways in which a continental surface can affect a planet's climate, including (but not limited to) surface albedo, topography, thermal inertia, surface roughness, etc. Depending on the nature of the land surface, these parameters could change significantly and thus alter the planet's climate \citep{Madden:2020}. 

The most extreme configuration in which continental surfaces play a major role is ``land planets''. Land planets, sometimes also known as dune planets, are desert rocky planets with limited surface water \citep{Abe:2011}. It is thought that this type of planet is one of the most probable (along with water-rich, ocean planets) around M-stars, as a result of formation and escape processes \citep{Tian:2015}. These types of planets have already been studied with 3D GCM simulations \citep{Abe:2011,Leconte:2013,Menou:2013,Yang:2014,Kodama:2019,Way2020} and specifically applied to the TRAPPIST-1 planets \citep{Wolf2017,Turbet:2018aa,Rushby:2020}. \cite{Rushby:2020} recently provided a detailed analysis of how surface type/composition may affect the climates of TRAPPIST-1 assuming they are land planets. This work was also described in the presentation given by Aomawa Shields.

While the study of continental surfaces of the Earth and other planets of the solar system (in particular Mars, characterized by its hyper-continental climate) is today our primary source of information on this matter, characterizing the climate of the planets of the TRAPPIST-1 system and other nearby planets (e.g., Proxima~b) may provide us with crucial data on how continental surfaces are operating on alien worlds.

\subsubsection{Ocean} \label{subsubsec:ocean}

Ocean modeling is often overlooked by the exoplanet community, largely due to the large computational expense associate with spinning up dynamic ocean models, coupled with the challenges of observing an ocean on another planet \citep[e.g.,][]{Robinson2010}. Future exoclimate simulations of terrestrial worlds benefit from the use of a dynamic ocean component \citep{Way2018,Yang2019a} coupled with a dynamic sea ice model.  It has been shown that sea ice drift can alter the habitable zone limits in various cases \citep{HuYang2014,Way2017,Way2018,DelGenio2019}.
Near the inner edge of the HZ, \cite{Leconte2018,Yang2019a,Salazar_2020} have shown that ocean heat transport is not always necessarily critical, especially when continents are present.  Warmer planets (T$_S$ $>$ 300 K) tend to have more  homogeneous surface temperatures, and thus ocean currents may not cause a meaningful net change in ocean-atmosphere heat exchanges.  \cite{Yang2019a} show this to be the case, and further state that ``...ocean dynamics have almost no effect on the observational thermal phase curves of planets near the inner edge of the habitable zone.'' These results suggest that future studies of the inner edge may devote computational resources to atmosphere-only processes such as clouds and radiation". However, \citet{Yang2019a} also argue that ocean heat transport is critical for the climate and observables for ``middle HZ'' planets in M-dwarf systems.  Still further, \cite{Way2018} demonstrated that planets with modern Earth-like land-sea masks show marked differences in mean surface temperature for fast rotators (spin less than 8 Earth sidereal days) versus slow rotators (see \citealt{Way2018}; Figure 2) in their inner edge of the habitable zone studies.

In addition, ocean composition---specifically, ocean salinity---may affect the fraction of habitable surface area and phase curves of middle and outer HZ worlds \citep{Cullum2016,DelGenio2018,Olson2020}. Salinity (dissolved salt content) is a first-order control on the density of seawater, and it further modulates the relationship between temperature and density. Salinity thus influences ocean stratification, circulation, and heat transport. At the same time, salinity depresses the freezing point of seawater, potentially limiting the formation of sea ice in salty oceans with consequences for surface albedo. In sum, saltier oceans tend to result in warmer climates. Models that neglect ocean dynamics cannot currently simulate salinity impacts on OHT, but may include freezing point depression. However, the relative contribution of heat transport vs. freezing point depression to climate warming, and how the balance of these effects may differ under different climate states is not well understood. It is thus unknown when/if simple hacks such as adjusting the freezing point of seawater in a model without a dynamic ocean is a reasonable strategy for simulating the climates of exoplanets with unknown ocean salinity or whether the likelihood that exo-oceans differ from Earth’s ocean with respect to salinity requires inclusion of a dynamic ocean. 

To facilitate future studies, the exoplanet modeling community should consider working more closely with oceanographers. For example, there is a large parameter space of ocean tidal dissipation to be explored that affects planetary rotation rates over time \citep[e.g.,][]{Green2019}. As mentioned above rotation rate has been demonstrated to affect climate. It must also be stressed that current GCMs used for exoplanetary studies have serious shortcomings in some cases. First, the putative thermodynamic oceans (also called q-flux \citep{Miller1983,Russell1985} as a heat source ``q'', whose values are generally specified by a control run, is prescribed to represent seasonal deep water exchange and horizontal ocean heat transport.) used in exoplanet GCMs have generally used zero horizontal heat transport, or highly simplified parameterizations \citep[e.g.,][]{Edson2012,Godolt2015,Kilic2017}. As well, current GCM dynamic ocean models are presently highly parameterized for modern day Earth since they are the children of Earth parent GCMs. For this reason it is important to engage more closely with the oceanography community to better parameterize the current suite of dynamic oceans used in exoplanetary GCMs.

For ocean planets that have a low density, the ocean depth can reach tens of or even hundreds of kilometers. At the bottom of the ocean, ice under high pressure may form. For the deep ocean, the equation of state for the seawater is required to be changed. Moreover, the ocean-bottom ice can influence the friction and the exchange of heat and materials between the ocean and the solid planet. Key questions may be answered using ocean GCMs: How deep are is ocean circulation (including both wind-driven and thermal-driven) and how does ocean circulation influence the concentrations of CO$_2$ and other greenhouse gases in the atmosphere \citep{Checlair:2019}.

Besides seawater oceans, another type that needs to be investigated are magma oceans. The lowest temperature of a magma ocean is about 1600 K. The density, viscosity, and diffusivity of the magma ocean are quite different from that of Earth's ocean. However, they may still of the same order. So, in this respect an Earth ocean GCM may be easily modified to simulate the  circulation of a magma ocean. But, a key process for the magma ocean is silicate (or other materials) precipitation in the ocean, which acts like vertical convection and can significantly influence the heat and mass transports in the ocean.

\subsection{Middle and Upper Atmosphere Processes} \label{subsec:midup}
Humans spend nearly their entire lives in Earth's troposphere, handling day to day local weather, and coping with climate change. Analogous stratospheric regions, and those above, constitute our best opportunities at characterizing an exoplanet's atmosphere and will likely, and should be, a primary focus in GCM development. The multitude of planetary GCMs that have been adapted for exoplanets have seen limited efforts incorporating middle atmosphere effects, especially coupling transport processes from other regions. Convection on Earth can lead to the production of high-altitude clouds (and  hazes) and being a source of atmospheric gravity waves. This represents a prime example of tropospheric-stratospheric coupling where momentum is transferred from the lower to the upper atmosphere. While there is work needed on features and processes that more directly influence the surface of exoplanets, further understanding and development of middle (and upper) atmospheric modeling in GCMs offers the greatest opportunity for scientific advancement in our interpretation of exoplanet observations from the next generation of telescopes.

\subsubsection{Non-equilibrium or non-conservative radiative and dynamical effects} \label{subsubsec:NLTE}
\paragraph{Gravity waves} 
The challenge of current exoplanet GCMs is to overcome the expense of running simulations with the necessary horizontal and vertical resolution over a range of pressures that adequately resolves processes operating over a variety of spatial and temporal timescales. But if our focus is data driven, we ought to look at the processes important in Earth's (and other worlds') middle atmospheres. Earth, Venus, and Titan, are all terrestrial worlds with significant super-rotation in their middle atmospheres. Venus' slow (and counter) rotation make it a hallmark case study for tidally locked exoplanets with substantial atmospheres. It has super-rotation in the equator that is likely driven by the Gierasch-Rossow-Williams (GRW) mechanism \citep[][]{Gierasch1975, RossowWilliams1979}, where planetary waves from high latitudes can transport angular momentum towards the equator and spin up super-rotating jets. Titan too has been observed to have significant variability in winds at different altitudes spanning from the stratosphere to the lower thermosphere and it too is a slow rotator at roughly 16 days. Earth has an alternating stratospheric jet oscillation, known as the Quasi-biennial Oscillation (QBO), which is the product of a complex interaction from a broad spectrum of waves. Stratospheric oscillations are also found to occur in giant planets of the solar system \citep{Fouc:08} and are a recent focus for climate modeling \citep{Cose:17,Bard:21}. The point is that their upper atmospheric dynamics are driven by waves not easily resolved and their effects are mostly missing from current exo-planet models while potentially having strong observational effects. 

Small-scale or high-frequency gravity waves have a large role in the upper atmospheric dynamics of terrestrial size planets. Not only are they an important source of momentum to drive the QBO, but they also affect mid-latitude jet streams, the semi-annual oscillation, and even the Brewer-Dobson circulation \citep{Butchart2014}. They are an efficient means of transporting energy across latitudes and altitudes and effectively redistribute energy, either mechanical or thermal away from its source to other areas until equilibrium or relaxed/balanced states are reached. Gravity waves try to keep the upper atmosphere of Earth in dynamical and radiative equilibrium by dispersing energy spatially over time. 
Gravity wave activity is also confirmed in Venus and Mars middle/upper atmosphere by several measurements and claimed to produce the observed variability in density, temperature and cloud structure \citep[e.g.,][]{Creasey2006, Garcia2009, Altieri2012}. 
Linear wave theory has allowed parameterizations of the effects from gravity waves and their breaking action in Earth based GCMs for some time. The typical approach is to assume some characteristics of the properties of the waves, amplitude or momentum flux, wavelength, phase speed, etc. that are inputs into the parameterization which is applied depending on the modeled local atmospheric environment - with a focus on horizontal and vertical wind shear and temperature gradients. 
Nevertheless, given the lack of systematic observations of gravity waves and the uncertainty of the source (for instance on Mars and Venus) which are necessary to constrain model parameters, our experience with different GCM configurations let us conclude that the total zonal wind (i.e. averaged for all local times) value is very sensitive to many GCM quirks. Zonal wind in the middle/upper atmosphere can be either positive or negative, producing different circulation regimes.

\paragraph{Non LTE effect in the upper atmosphere}
The upper atmospheres of terrestrial-like planets in our Solar System are similar in terms of physical processes, in spite of important differences in temperature, density and composition \citep{Gladstone2002}. A basic property of these upper layers is the low gas density that is also responsible for situations of breakdown of Local Thermodynamic Equilibrium (LTE), specific for each molecular species and each vibrational transition. These non-LTE effects result in populations of the molecular energy states not dictated by Boltzman statistics at the local kinetic temperature and occur when molecular collisions are so infrequent that other processes (e.g., radiative transfer) become important for the determination of those states' number population  \citep{LopezPuertasTaylor2001}. In terrestrial planets, and for the main molecules and infrared emissions, those layers usually correspond to their mesosphere and thermosphere. 
At pressure layers above about 10$^{-5}$ mbar, solar EUV heating and thermal conduction are the main processes controlling the energy balance, while in the mesosphere (on terrestrial planets between 1 and 10$^{-5}$ mbar, approximately), absorption and emissions by atmospheric molecules with active ro-vibrational bands in the IR usually play a crucial role on the thermal structure \citep{Gladstone2002}. 

These non-LTE processes have to be considered when interpreting strongly irradiating exoplanets. CO$_2$ and CO non-LTE fluorescence is common in telluric atmospheres, but CO emission has also been detected in Neptune \citep{Fletcher2010}. Furthermore, non-LTE radiative transfer modelling helped to explain unexpected observed features around 3.3 um on the hot-Jupiter HD 189733b from ground measurements, reported to be CH$_4$ non-LTE emission \citep{Swain2010,Waldmann2012}.
Future detection by JWST, LUVOIR, together with ground-based measurements of upper atmosphere of hot-Jupiters by IR spectrographs (CRIRES/VLT, METIS/E-LTE) will make it possible to test composition and temperature models of warm and hot Jupiters, However, it is still challenging for terrestrial exoplanets.

Due to its expected significant effect on exoplanet atmospheric characterization, the improvement of the mid and upper atmosphere processes in exoplanet atmospheric models should therefore be a priority in the era of JWST and ELTs.


\subsubsection{Photochemistry in 3D atmospheres} \label{subsec:photo}

The transport of photochemically produced gaseous species and hazes could have a significant impact on the characterization of exoplanet atmospheres \citep{Carone2018}. Different circulation regimes can result in different global distributions of important atmospheric species such as ozone \citep{Yates_2020,Chen2019}.
The community should first think about what kind of composition is the most interesting and imperative for near-term observations (e.g.,  exo-Venuses). Very large uncertainties remain for many deposition fluxes. For instance CO and relatively minor tracers like Cl have an important catalytic activity for both Venus and Mars but are not accurately constrained. Along a similar vein, the surface emission fluxes of various biogenic compounds such as DMS are unconstrained, but different assumptions can dramatically effect their resultant global distributions \citep{Chen2018}.

Exoplanet observations of many regimes show that clouds and hazes significantly affect planetary spectra. Current GCMs provide the spatial mapping of water clouds, which can have strong effects on transmission spectra and thermal phase curves \citep{Wolf2019}. However, the majority of GCMs do not include self consistent treatment of photochemical haze. As an integral part of climate modeling, aerosols need to be included 
- either from the ground up (production rates from chemical networks), or decoupled  (aerosols and photochemical hazes would be separated). For instance, in modeling Titan (and Titan-like exoplanets) \citep{Lora2018}, the production rate is fixed to reproduce observations, and then the photochemistry is separated. For exoplanets, this would required the use of very detailed models to capture monomer formation self-consistently, that will be used in a simple photochemical model favoring the approach of a lower-resolution model but with a higher-complexity component. As a follow-up to \citet{Chen2019}, one possibility is through adapting CARMA \citep{Larson2015}, a state-of-the-art microphysical model that can be used here to simulate the evolution of hazes.  Note that CARMA is already coupled to ExoCAM with a nominal fractal aggregate haze model \citep{Wolf+Toon2010}, with funded plans to use it for studying hazy iterations of the habitable zone planet TOI-700 d.  Haze production rates will be sourced from off-line 1D Atmos calculations (e.g., \citet{Arney2017}.
Lastly, the 3D temperature structure of a planet could also be important for the chemistry itself (exo/endo-thermal reactions), even in absence of photochemistry,

It has to be noted that for exoplanet photochemistry, the use of pre-calculated photolysis rate tables is much too specific. A rate table of reasonable size is unlikely to be generic. For an atmosphere for which we do not know the profile of the species we cannot build a coherent table. It is necessary to go through an online calculation of the photolysis rates which leads to a reconsideration of the radiative transfer within the GCM. Since most GCMs use the correlated-k method it cannot be used for photolysis calculations. In short, it is necessary to calculate the photolysis online. However, simulating the photochemistry and chemistry with 3D models is computationally very expensive. It may require prior work to reduce the number of gaseous families, which would depend on bulk composition, temperature and instellation. Such an approach has been started by Olivia Venot's group but it is uncertain how 3D transport will affect this optimization which is performed in 1D. 

Another caveat is that the majority of terrestrial 3D chemical models are restricted to present-day Earth compositions. This is due to the fact that these models largely inherited components from their model supersets originally developed for Earth-based research. For instance, \citet{Chen2019} adapted the National Center for Atmospheric Research model \citep{Marsh2013} by deploying subroutines from ExoCAM with the Whole Atmosphere Commuity Climate Model (WACCM). Thus another goal is to extend 3D photochemistry models non-oxygenated reducing and weakly-oxidized (H$_2$-, N$_2$- and CO$_2$-rich) atmospheres. Such anoxic conditions dominate the atmospheric evolution histories of Earth, Mars, Venus, and Titan. This suggests that anoxic atmospheres are the ``default" state of a planet's eventual fate in the absence of biological activity. As shown by previous 1D work \citep{Lincowski2018}, assessing non-Earth-similar atmospheres is important to gauge the detectability of photochemical byproducts on the myriad of potential rocky exoplanet compositions.

While out of the scope of this report, it is important to mention that star-planet interaction is fundamental to understand a range of key processes (stellar-sourced charged particles, EUV heating, photodissociation, and photoionization, etc) that could shape the upper layers of rocky exoplanet atmospheres. Collaboration with stellar physicists and observers will be crucial.

\subsection{Aerosols in exoplanet GCMs}\label{subsec:aerosol}
\subsubsection{Condensible gases in GCM simulation for exoplanets} 

Characterization of non-water condensables is extremely important for understanding both the atmosphere and surface processes of other worlds. 

In the low-temperature range, gases such as CO$_2$, H$_2$S, SO$_2$ can condense \citep{Fray:2009} for planets with hydrogen-dominated atmospheres and volcanic activity; while CH$_4$, NH$_3$, and N$_2$ are expected or found for worlds like Venus, Titan, Pluto and other Solar System moons. This condensation can have large consequences on the planet's atmosphere by removing greenhouse gases, forming clouds and modifying surface albedo. For instance, \cite{Turbet2017} showed that CO$_2$ condensation can strongly reduce the deglaciation of terrestrial planets as CO$_2$ condensation leads to accumulation of surface CO$_2$ ice that can get permanently trapped under water ice.
In the high-temperature range, a variety of condensables would potentially be observable by JWST or future extremely large telescopes (ELTs) for highly irradiated exoplanets, but data on properties (e.g., microphysical properties) of these condensable species is sorely lacking. 

Spectral information would be needed from laboratory experiments and missions regarding optical properties of exotic clouds and albedo properties for non-water condensables, to improve model response to a broad variety of scenarios.  

In addition to acquiring new data, model developments are needed within GCMs in order to include condensable species other than water and to precipitate condensables with appropriate albedo, grain-size properties. In this area, the LMD-G GCM already has extensive capabilities for CO$_2$ condensation on Mars \citep[e.g.,][]{Forget98}, early Mars \citep[e.g.,][]{Forget:2013} and exoplanets \citep[e.g.,][]{Wordsworth:2011,Turbet2017}, for N$_2$ condensation on early Titan \citep[e.g.,][]{Charnay:2014}, N$_2$, CH$_4$ and CO condensation on Pluto \citep[e.g.,][]{Forget:2017}.

Overall, much more data is needed to make significant progress on the question of condensible species in exoplanet's atmospheres and surfaces (in particular, for high temperature condensables), along with potentially complicated and time-consuming model development. Initial work coupling various schemes with multiple condensate species has been done for hot Jupiters \citep[e.g.,][]{lee_2016,Lines_2018,Lines_2018b,Lines_2019}.

\subsubsection{Impacts of aerosol microphysics in GCM simulations and simulated spectra} \label{subsec:micro}
Aerosols are present in every atmosphere of our solar system planets and moons. Clouds have also been observed in exoplanet's atmospheres such as the super-Earth GJ-1214b \citep{Kreidberg2014}, the gaseous
giant WASP-12b \citep{Wakeford2017}, and WASP-31b \citep{Sing2016}. Hazes have been observed on WASP-6b \citep{Nikolov2015} and HAT-P-12b \citep{Sing2016}.  They have not been observed for terrestrial size exoplanets yet but simulations using GCMs and PSG have shown that they dramatically flatten the transmission spectrum preventing an exhaustive atmospheric characterization from space observatories \citep{Fauchez:2019,Komacek2020,Suissa2020}.

However, the detailed aerosol microphysics is uncertain for exoplanet atmospheres. Yet, changes in the microphysical and optical properties can have a very large impact on the climate simulation and  simulated spectra. To improve our understanding of aerosol properties in exoplanet atmospheres we need data, from the lab but also from JWST and ARIEL. If a statistical study is performed on such data it may allow us to discriminate between cloud particles and hazes that differ in terms of size and microphysical/optical properties. Also linear polarization is a powerful tool with which to retrieve cloud microphysical/optical properties as is performed for Earth using, for instance, POLDER/PARASOL data \citep{Goloub2000}. Finally, we have to improve our connection with other communities such as the Earth scientists, paleoclimatologists, and Solar System planetary scientists to better share data (remote sensing + in situ) and methods of applying data to exoplanet atmospheres.

In addition to data modeling studies can also be very helpful. For instance, a sensitivity study on cloud microphysics can be performed by varying cloud particle size and amount of cloud condensation nuclei (CCN) to simulate their impact on simulated spectra. Also, modeling work could allow to better understand how spatial heterogeneity (horizontal and vertical resolution, cloud gap fraction, overlap) affect both transmission and reflection spectra. These improvements may necessitate the inclusion of biological coupling for both haze (microbial, plant) and fires and therefore to develop computational physics packages, chemistry packages, and increased capabilities for capturing spatial and temporal heterogeneity.

More direct collaboration needs to be done with the climate modelers and observation simulators to bring consistency between assumptions about radiative transfer. 
More observations of Earth transmission spectra are therefore needed, for example the Atmospheric Chemistry Experiment - Fourier Transform Experiment (ACE-FTS) onboard the SCISAT-1 satellite provides observation in the 2.2$\mu m$ - 13.3$\mu m$ window. This emphasizes the benefit of Earth and Planetary Science synergies.

\subsection{Synergy between EBMs, 1D radiative-convective and  photochemical models with GCMs} \label{subsec:syn}

\subsubsection{EBMs in the THAI workshop}
The HEXTOR energy balance model (EBM) was used to conduct the THAI scenarios, using both a latitudinal and longitudinal mode \citep{haqqmisra2021}. HEXTOR is a one-dimensional EBM based on the model by \citet{Williams1997}. The model is typically run in a latitudinal mode, which reproduces Earth's mean annual climate. The model can also be used to explore changes in Earth's climate due to past and future orbital variations and possible feedback from anthropogenic forcing \citep{haqq2014damping}. Prior versions of this model have represented radiative transfer with a basic linear relationship \citep{haqq2014damping} or with a polynomial fit of 1D radiative-convective climate calculations \citep{Williams1997,HaqqMisra2016,batalha2016climate,hayworth2020warming}. The current version of HEXTOR attempts to improve the accuracy of the radiative transfer in the model by using a lookup table, which conducts a nearest-neighbor interpolation for OLR and albedo using a database containing thousands of 1D radiative-convective climate calculations. This provides an advantage in accuracy at the cost of added computational expense. During the THAI workshop, Dr. Haqq-Misra showed that HEXTOR in a latitudinal configuration either underestimates or overestimates the global average temperature in the THAI simulations, because the hemispheric differences between the day and night sides cannot be represented with a single dimension in latitude; however, he also showed that HEXTOR can also be configured as a longitudinal EBM through a coordinate transformation, which places the substellar point at the north pole and allows the day-to-night side contrast to be represented more accurately \citep{fortney2010transmission,koll2015deciphering,Checlair:2017,haqqmisra2021}. Longitudinal EBMs, either along the equator like HEXTOR or with full latitude-longitude resolution, can provide constraints on climate across broad parameter spaces or for long time integrations, which can be useful in identifying specific problems to study further with GCMs.

VPLanet \citep{Barnes2020} includes an EBM called POISE (Planetary Orbit-Influenced Simple EBM), a one-dimensional seasonal EBM that reproduces Earth's annual climate as well as its Milankovitch cycles \citep[see][]{North1979, HuybersTziperman08,Deitrick2018b}. Though the model lacks a true longitudinal dimension, each latitude is divided into a land portion and a water portion, with distinct heat capacities and albedos, and heat is allowed to flow between them. Ice can accumulate on land at a constant rate when temperatures are below 0$^{\circ}$ C, while melting/ablation occurs when ice is present and temperatures are above 0$^{\circ}$ C. Sea ice forms when a latitude's temperature drops below $-2^\circ$ C (accounting for salinity), and melts when higher. To account for ice sheet flow, bedrock depression, lithospheric rebound, and ice sheet height, they employ the formulations from \cite{HuybersTziperman08}. The bedrock depresses and rebounds locally in response to the changing weight of ice above, always seeking isostatic equilibrium. POISE is thus a self-consistent model for ice sheet growth and retreat due to instantaneous stellar radiative forcing, orbital elements, and rotational angular momentum. 

\subsubsection{1D radiative-convective and photochemical models in the THAI workshop}
1D radiative-convective climate and photochemical models are one dimensional models representing a vertical atmospheric column assuming plane-parallel waves in hydrostatic equilibrium. In the photochemical models, the vertical transport takes into account molecular and eddy diffusion and are able to represent a complex photochemistry. 
1D models have been widely used by the community to determine the edges of the habitable zone \citep{Kopparapu2013}, to study the ancient Earth \citep{Arney2016,Arney2017} and various exoplanets \citep{Lincowski2018,Meadows2018}. In this workshop, THAI simulations with the Atmos 1D model \citep{Wunderlich2020} were presented by Andrew Lincowski following a two-column approach. Dr. Lincowski has shown that two 1D radiative-convective atmospheric columns are able to reproduced the day-night temperature contrast simulated by GCMs while keeping an advantage in term of computational time (Lincowski et al., in preparation in this focus issue).

\subsubsection{Synergy between GCMs, EBMs and 1D models}

GCMs are very complex models that require significant time to converge. Lower dimensional models such as EBMs or 1D radiative-convective climate and photochemical models, while  ideal to explore large parameter sweeps, lack a representation of atmospheric dynamics, surface heterogeneity and clouds. The computational efficiency of EBMs enable them to simulate climates on much longer timescales of thousands or millions of years to explore orbital and rotational effects on climate \citep[e.g.,][]{Spiegel2009,Deitrick2018b}.
EBMs are typically 1D in latitude and solve a single partial-differential equation for surface temperature. Temperature then depends on incoming stellar flux (instellation), heat diffusion, albedo, and the outgoing longwave radiation (OLR). The OLR and albedo are parameterized with simple formulations \citep{North1979,Spiegel2009,Rose2017,Palubski2020}, though several studies have made advancements by fitting polynomials to radiative-convective models \citep{Williams1997,HaqqMisra2016}. The chief challenge of these models comes from the parameterization of atmospheric dynamics in terms of a heat diffusion term and accuracy in parameterizing the radiative transfer. For this reason, synergy with GCMs is necessary to ensure some measure of accuracy and predictive power from EBMs.

1D EBMs and radiative-convective climate models coupled to photochemical models can explore a very large parameter space (i.e., star and planets properties, instellation, rotation and orbital periods, eccentricity, atmospheric properties, etc.) and can identify key points of interest in the parameter space that 3D models can then investigate. 
For instance, 1D models can be used to determine the likely chemical state as input to a GCM. GCMs can also be used to determine cloud coverage percentage and dynamical model as input to 1 and 1.5-D models. GCMs can be run with simple tracer chemistry with haze precursors and 1D photochemical models can be used to figure out what happens next with haze formation/chemistry etc. Using both 1D and 3D models simultaneously would allow one to get a more complete picture of chemistry, clouds and observables.

It has also been highlighted during the THAI workshop that interactions between the atmosphere and interior of terrestrial planets require more attention that currently given. This would require improving the collaboration with geologists/geophysicists. Such coupling should probably first be developed in 1D following a ``Planet evolution model" approach based on an asynchronous coupling employing a mixture of (short term) climate calculations and long term simulations (for glaciers, but also longer processes as well).

Finally, it is important to predict in advance, with a hierarchy of models, what we might see and have the models ready to interpret the data. Ideally, upstream modeling work should not be constrained by anticipated observational sensitivity. 

\section{The Future of Exoplanet GCMs, Results of the Pre-Workshop Survey} \label{sec:survey}

Several weeks prior to the workshop, an online survey was sent to all THAI participants to poll their opinion on what the field of exoplanet GCMs might look like in the coming decades. The aim of this exercise was mainly to highlight key modeling developments that need to be pushed by the community to move the field forward in the best possible directions.

A total of 35 participants completed the online survey. Participants have different levels of career advancement (3 undergraduate students, 4 graduate students, 13 early-career scientists, 12 mid-career scientists, and 3 senior scientists) and work in several continents (17 in Europe, 14 in North America, 5 in Asia, and 1 in Oceania). The survey consisted of a dozen questions, the main results of which are summarized below.\\

\begin{figure}
 \includegraphics[width=9cm]{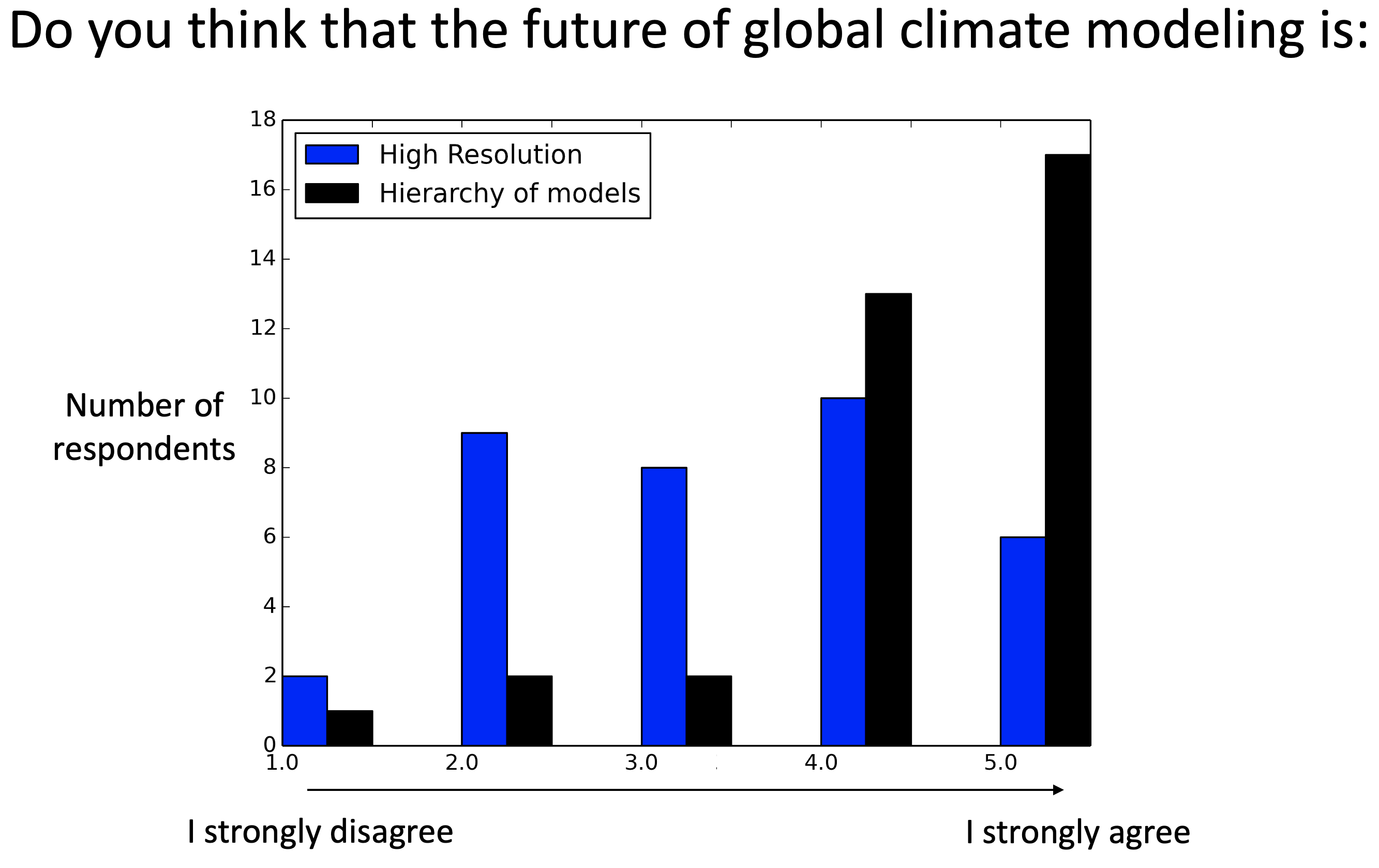}
 \caption{Results of the first item of the survey: ``Do you think that the future of global climate modeling is?" (1) First possibility (in blue): High (spatial) resolution thanks to increased computing resources?
(e.g., to explicitly simulate convection processes directly in GCMs) and (2) Second possibility (in black): Using a hierarchy of models ranging from very fine resolutions to global scale? (e.g., to simulate explicitly convection in an idealized box to derive subgrid scale parameterizations for GCMs). }
 \label{Q1_survey}
\end{figure}

\textbf{(1) High resolution simulations: global or local?}
With the increase in available computing resources, and the need to simulate atmospheric processes such as convection and clouds without using empirical parameterizations, high spatial resolution seems to be an attractive development pathway of the exoplanet GCM field for the coming decades. It is in fact one of the main directions of development in the modeling of the future of Earth's climate \citep{Stevens2019}. We asked the survey participants if they thought that the future of very high spatial resolution simulations for exoplanets was on the side of global or local simulations (i.e. simulations performed on a local grid and then used to derive parameterizations of subgrid processes to be used in low spatial resolution GCM simulations). The results, which are presented in Fig.~\ref{Q1_survey}, show that most respondents believe that the hierarchical approach (local high-resolution simulations to derive subgrid parameterizations for low resolution GCM simulations) is the most promising for the field. It should be noted that several recent works on exoplanet atmospheric modeling go in this direction \citep{Zhang:2017,Koll:2017,Lefevre:2019,Sergeev2020}. \\

\begin{figure}
 \includegraphics[width=9cm]{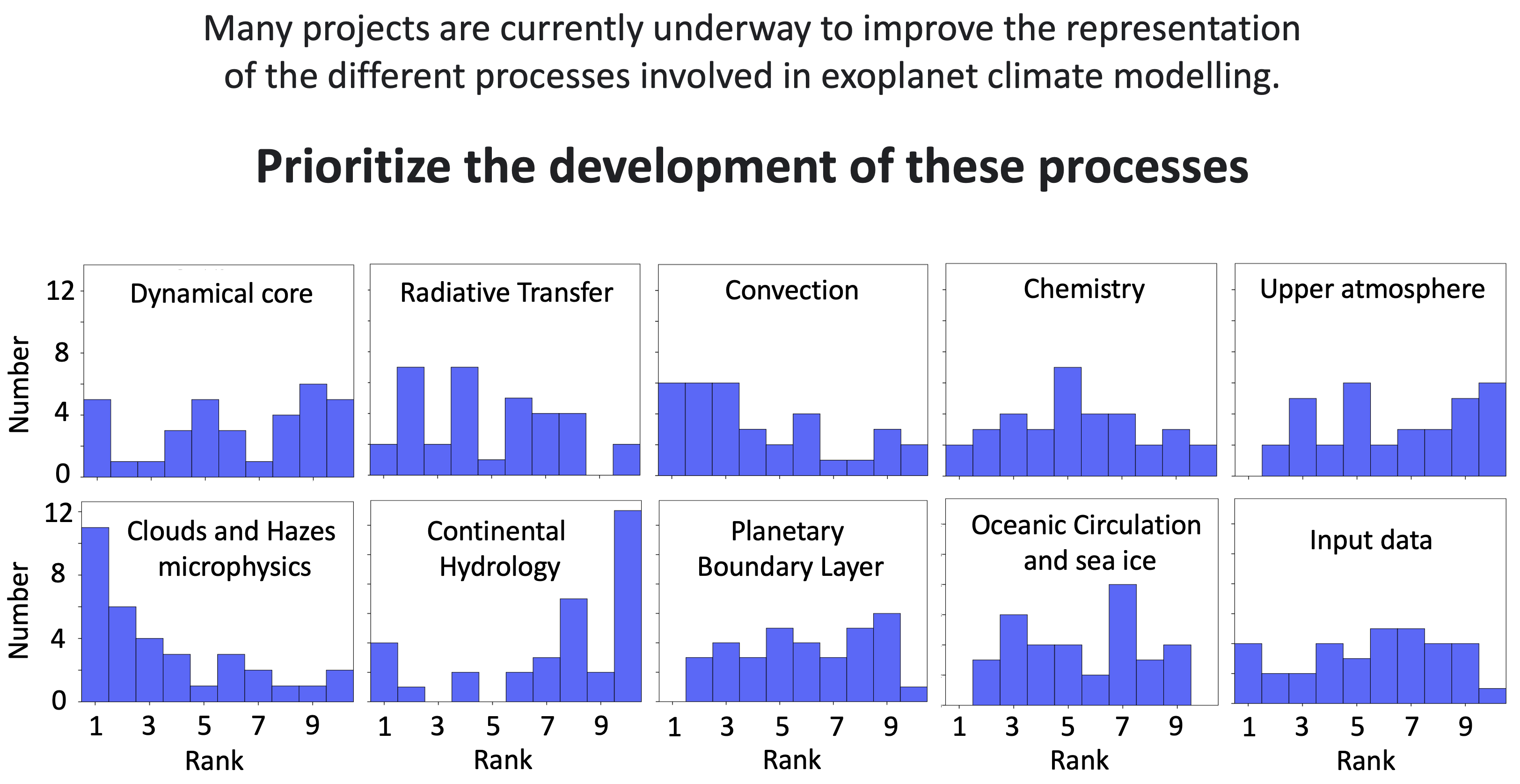}
 \caption{Results of the second item of the survey: ``Many projects are currently underway to improve the representation of the different processes involved in exoplanet climate modelling. Can you prioritize the development of these processes in the order you consider appropriate below? (1 is highest priority ; 10 is lowest priority) ------------------- For example, choose 1 for convection (1st priority) ; 2 for continental hydrology (2nd priority) ; 3 for cloud/hazes microphysics (3rd priority) ... until dynamical core (10th and last priority)" }
 \label{Q2_survey}
\end{figure}

\textbf{(2) Most important processes to be modeled in fully coupled 3D GCMs}
As more computing resources become available, it is becoming increasingly possible to build fully coupled 3D GCMs, i.e. GCMs that include all processes at play (chemistry, aerosols, oceans, glaciers, etc.) in/on a planetary atmosphere. It is by combining all these processes at the same time -- in the same way that it is done for fully coupled Earth GCMs \citep{Sellar:2019} -- that it will be possible to build virtual planetary atmospheres that are more and more realistic and therefore able to interpret the observations. This is in this context we asked the survey participants to prioritize the processes for which it is most important today to focus our efforts. The results, which are presented in Fig.~\ref{Q2_survey}, show that most respondents ranked clouds/hazes and convection as the first and second most important processes for the field to focus on. This is most likely because clouds/hazes (and moist convection, which leads to cloud formation) have been identified as the most serious limit for probing the composition of exoplanetary atmospheres, in particular using the transit spectroscopy technique \citep{Fauchez:2019,Komacek2020}. This interpretation is also reflected in the results of the open-ended question ``According to you, which developments should be prioritized to connect GCM models to ongoing and future observations of exoplanets?". The vast majority of those who answered this question did indeed mention clouds as the top priority for modeling efforts.\\

\begin{figure}
 \includegraphics[width=9cm]{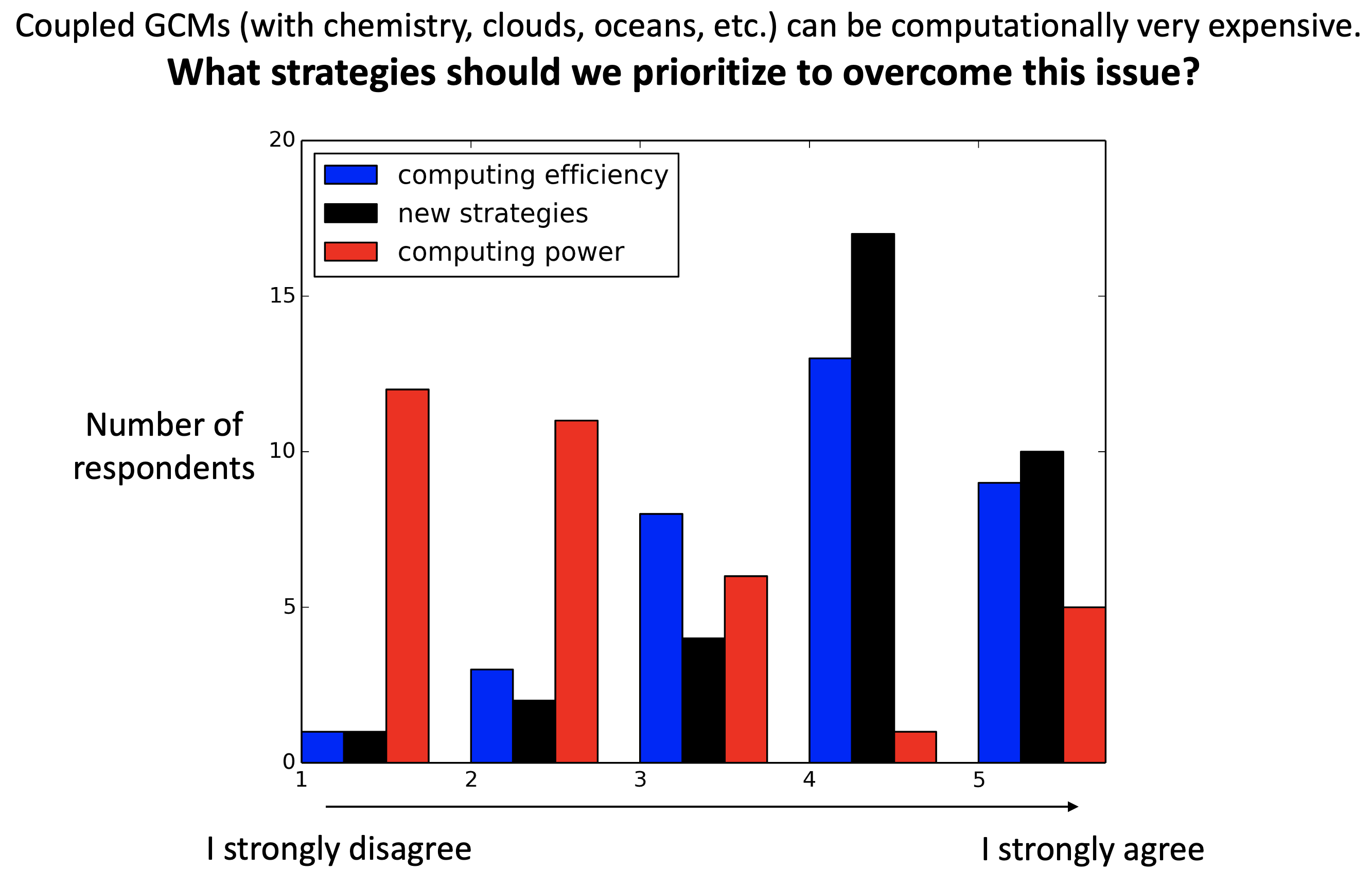}
 \caption{Results of the third item of the survey: ``Coupled GCMs (with chemistry, clouds, oceans, etc.) can be computationally very expensive. What strategies should we prioritize to overcome this issue? (1) First possibility (in blue): Improving GCM codes to make them less resource-intensive. (2) Second possibility (in black): Exploring new strategies (e.g., asynchronous coupling, convergent suite of simulations, etc.) to accelerate the convergence of GCM simulations. (3) Third possibility (in red): Thanks to Moore's law, this will not be an issue anymore in the future." }
 \label{Q3_survey}
\end{figure}

\textbf{(3) Best strategies to limit the computing time needed to perform fully coupled 3D GCM simulations}
Despite the increase in available computing resources, some  atmospheric and/or surface processes can be extremely costly in computing time. We thus asked the survey participants what they felt were the best strategies to address this issue. The results are presented in Fig.~\ref{Q3_survey}. Most respondents believe that the increase in computing resources will not be sufficient to address the issue, and that instead efforts should be put on improving the efficiency of numerical codes as well as on developing new strategies to accelerate the convergence of models.\\

\textbf{(4) Computing language}

 \begin{figure}
 \includegraphics[width=9cm]{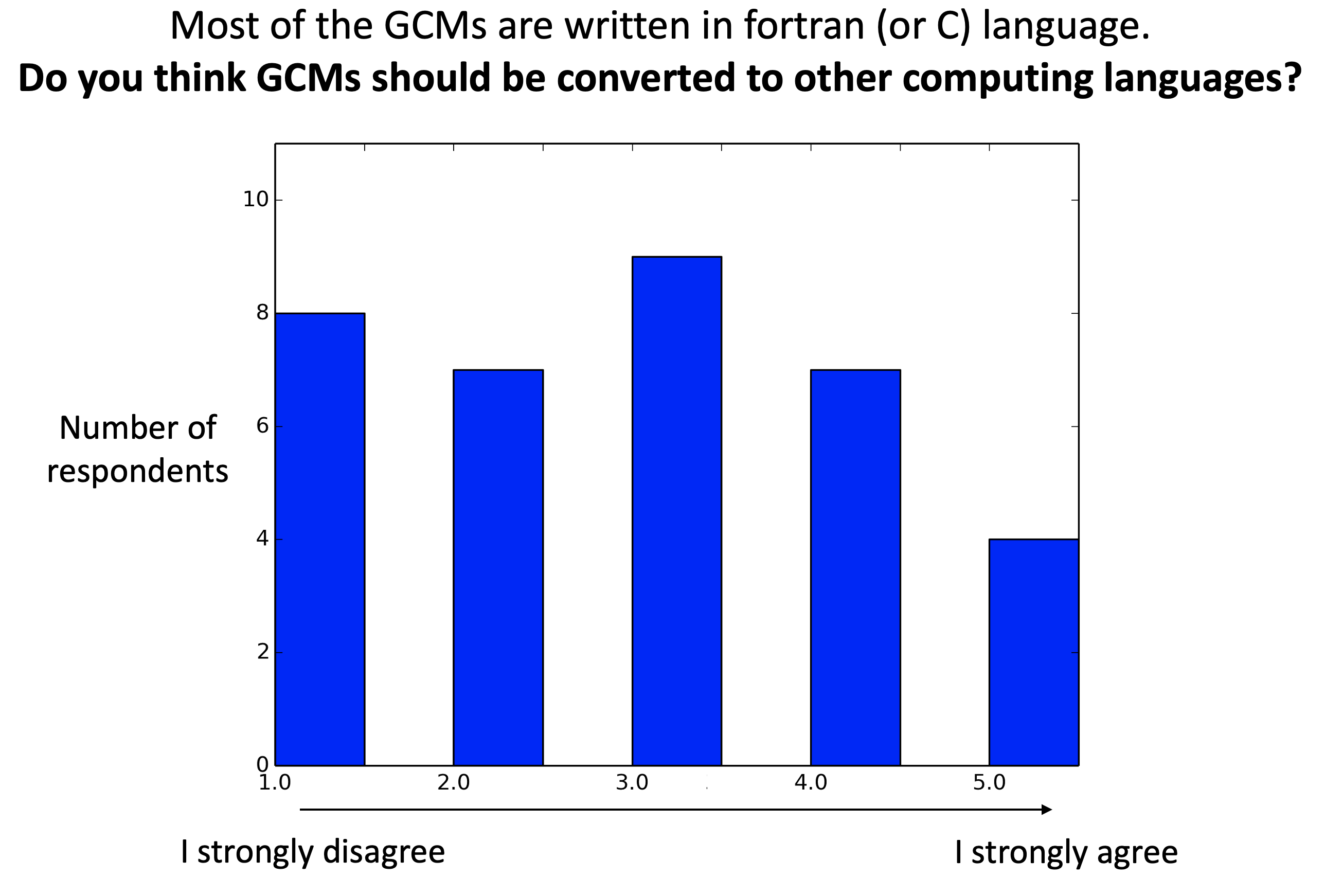}
 \caption{Results of the fourth item of the survey: ``Most of the GCMs are written in Fortran (or C) language. Do you think GCMs should be converted to other computing languages?" }
 \label{Q4_survey}
\end{figure}

The effectiveness of GCMs as well as their usability and their ability to evolve over time depends on the programming language in which they are written. Most GCM codes are mainly written in Fortran -- the oldest high-level programming language --- and one can thus wonder if these codes require to be converted to a more modern language, such as Python. However, the time require to convert these codes would be tremendous and Fortran remains a very fast language to perform the GCM calculations. We therefore keep using them as legacy codes.     

We therefore asked the survey participants if they believe that GCM codes should be converted from Fortran to modern languages (Fig.~\ref{Q4_survey}). Opinions were very divided, with a slight prevalence for negative answers. This issue has also been the subject of intense debate during the third day of the workshop.

Among the disadvantages of Fortran that have been put forward: 
\begin{itemize}
\item Fortran is difficult to handle for new generations of students (accustomed to other modern object oriented programming languages e.g., Python). This may highly impact the attractiveness of the field, with a risk that these students and, more generally, scientific and engineering developers may turn away and/or lose the skills for sophisticated computer development in Fortran.
\item The community of developers of modern languages (e.g., Python) is now much wider, and therefore there are many more libraries and contents that GCM codes could make use of.
\end{itemize}

And the responses from critics:
\begin{itemize}

\item Once students know one programming language, they can in principle easily adapt to other languages. 
\item Most GCM codes are several hundred thousand lines long, so in practice it is an excessive amount of work to convert a GCM code into another computer language.
\item Which language to choose for converting GCM codes? Python? C? How do you know if these languages will still be widely used 5, 10, 30 years from now?
\item Fortran is a very efficient (and evolving) programming language, e.g., last version if Fortran 2018. A first (reasonable) alternative is therefore to modernize the GCM codes to the most recent versions of Fortran. Fortran compilers are also highly optimized and fast.
\item Finally, a compromise could be found in using Python (or a graphic user interface (GUI)) as a wrapper to run a GCM for which the core code would be in Fortran. Note that the UM GCM already uses such GUI, a but it requires additional resources and funding to maintain and update it.

\end{itemize}
To summarize, the fact that Fortran is used for GCMs is historical, but continues to be justified because it is a compiled language that has evolved to offer high performance, in particular for parallel operations  on multicore or massively parallel environments. Alternative compiled languages are C or C++. Nevertheless, Python is currently the language growing in popularity to write scientific code in spite of the fact that is it is not a compiled language and thus much slower than Fortran, for instance. The runtime performance of Python can be improved by using pre-compiled libraries (e.g., numba or NumPy), but it has not yet been used to develop a GCM. \\

\textbf{(5) Machine learning}

 \begin{figure}
 \includegraphics[width=9cm]{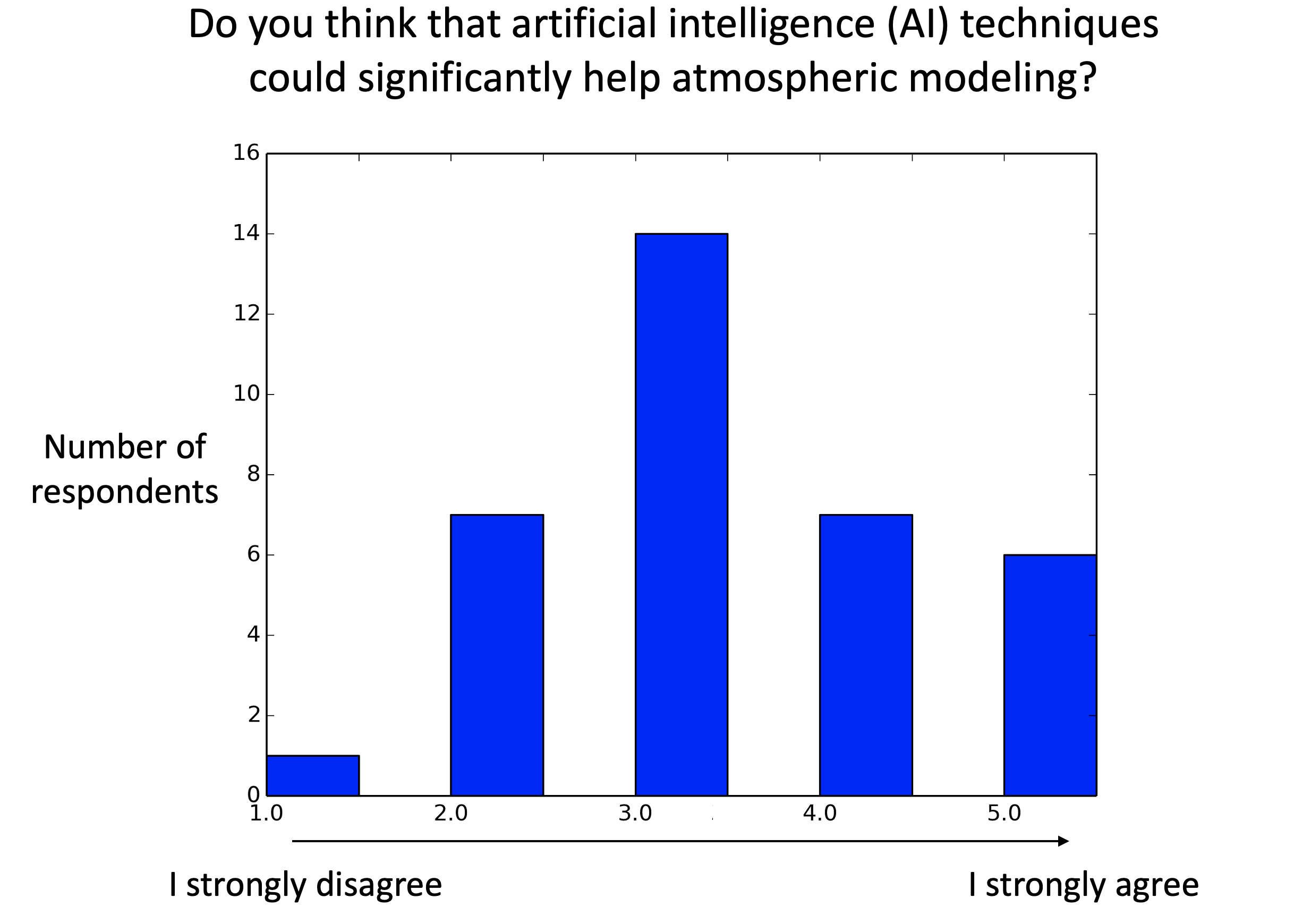}
 \caption{Results of the fifth item of the survey: ``Do you think that artificial intelligence (AI) techniques could significantly help atmospheric modeling?" }
 \label{Q5_survey}
\end{figure}

Machine learning techniques are on the verge of revolutionizing many fields of science, including astrophysics \citep[e.g.,][]{Way2012,Ivezic2019} and exoplanets \citep[e.g.,][]{Shallue2018,Armstrong2020}. We thus asked the survey participants if they believe that Machine Learning (ML) /Artificial Intelligence (AI) techniques could also significantly help atmospheric modeling and if so how. The results are presented in Fig.~\ref{Q5_survey}. Opinions are again very divided, but with a significant peak for people with no opinions. This is most likely symptomatic of the fact that the use of ML techniques is a topic that has been very little discussed in the (exoplanet) atmospheric modeling community to date. Some survey participants mentioned that ML techniques can be used to derive better sub-grid scale parameterization, e.g., of convection. It is an avenue being currently explored for the modeling of the Earth's climate (see e.g., \citealt{Rasp:2018}). These ML techniques could also prove to be a promising way to connect local 3D high-resolution cloud resolving models with 3D low resolution GCMs, in line with the first point of the survey.\\

\textbf{(6) Environmental impact of numerical simulations}

 \begin{figure}
 \includegraphics[width=9cm]{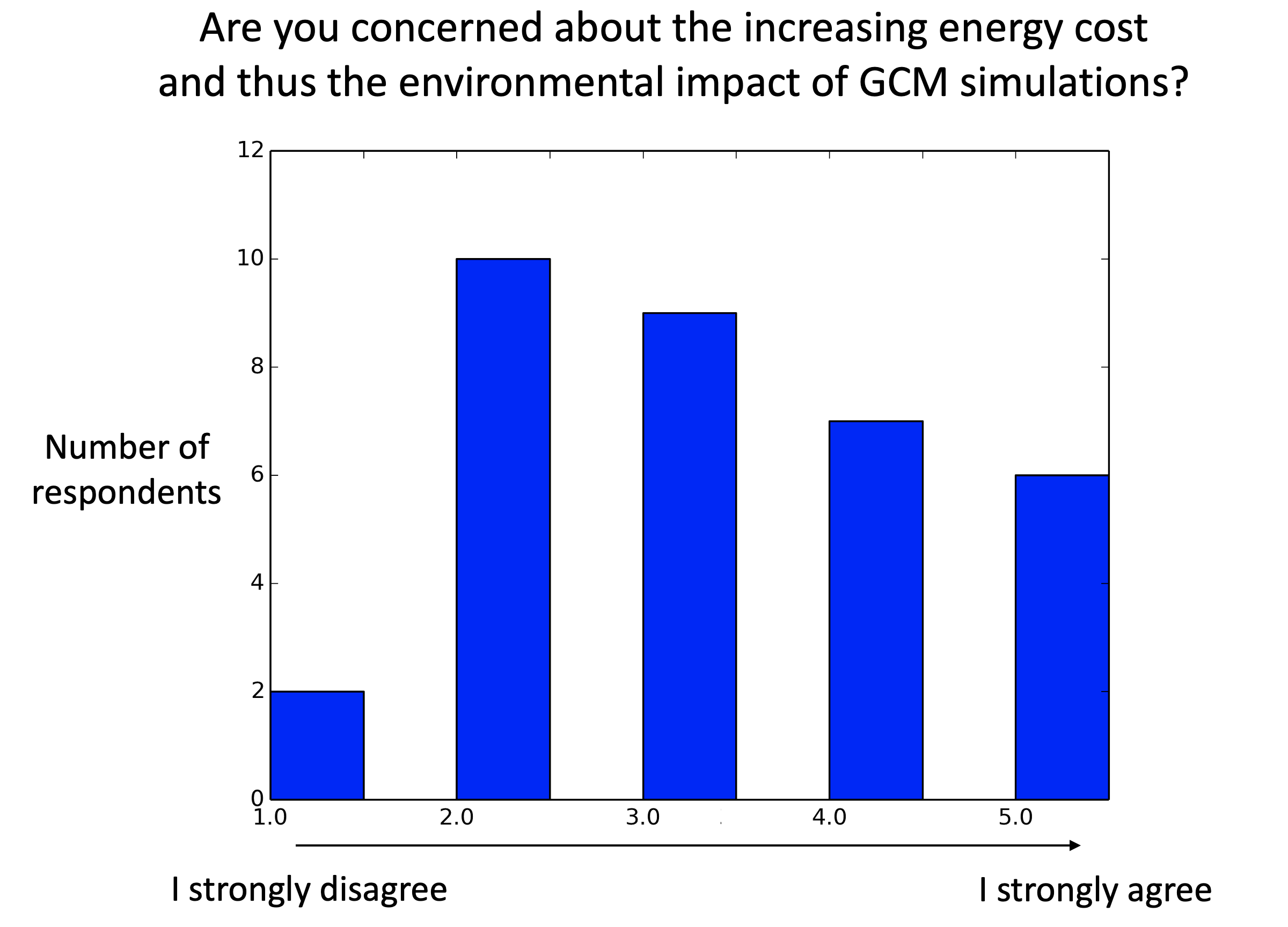}
 \caption{Results of the sixth item of the survey: ``Are you concerned about the increasing energy cost and thus the environmental impact of GCM simulations?" }
 \label{Q6_survey}
\end{figure}

Today's and especially tomorrow's GCM simulations (with the increase in both the resolution and number of physical and chemical processes taken into account) are and are likely to be very energy-consuming, with a potentially high environmental footprint (greenhouse gas emissions, rare-earth metal mining, etc.). We thus asked the survey participants if they were concerned about the increasing energy cost and thus the environmental impact of GCM simulations. The results are presented in Fig.~\ref{Q6_survey}. Opinions, that are again quite divided, were the subject of debate on the last day of the workshop.

One of the preliminary proposals that emerged from this discussion is to make the environmental impact of our work more transparent, for example by stating in our publications the amount of greenhouse gases (e.g., in CO$_2$ tons equivalent) that were emitted for the study. As this carbon footprint can vary by several orders of magnitude from one country to another (depending on the carbonation level of the electricity network), from one GCM to another, from one parameterization used to another, from low to high resolutions, or depending on the number of simulations performed, it is very difficult to know the emissions associated with each study. More transparency on this subject would raise the community's awareness and could ultimately contribute to impacting environmental policy decisions (e.g., at the level of researchers, so that they make the most intelligent use of available resources to avoid waste; at the level of the University in the choice of computing equipment, energy source of the cluster; at the national/international level, to encourage the decarbonization of the electricity networks).

It has also been mentioned that carbon offset strategies could be budgeted during proposal submission. However, the efficiency of carbon offset projects (including tree planting) is highly debated today \citep[e.g.,][]{Gates2021}. Finally, it has to be noted that short and small workshops such as the THAI workshop are very well suited for remote solutions and would help to mitigate the research laboratories carbon footprint release by flying to meetings.

This workshop report therefore recommends to GCM users to systematically disclose the amount of CO$_2$ released by running computer simulations and eventually consider a carbon mitigation plan.

While it has not been actively discussed during the THAI workshop, it is important to mention here the access of GCM data post publications. Discussions among co-authors generally agree that GCM data should be made available post publication, when possible. However, the amount of GCM data can be very large which may lead to additional fees to store them on disks and/or clouds beyond the limit that is usually allowed for free. It has also been discussed that it is actually quite rare that data from a published study are effectively downloaded and used. Therefore, the ratio benefit to cost of systematically making available GCM data may not always be relevant.
Also, some models are inherently proprietary and serve the community better that way than if they would become open source. Indeed, it requires a lot of resources and personnel to keep a large and complex code at the forefront of its field. This is the case for instance of the UM  owned by the UK Met Office. The proprietary license, however, does not prevent sharing output data and configuration files, which is the case for UM’s contribution to THAI.

\section{Creating a Diverse and Inclusive International Community in the Exoplanet GCM field} \label{sec:div}
The workshop also included discussions about taking concrete action to improve aspects of diversity, inclusion, equity, belonging, and justice that will have long term implications. The workshop organizers decided to include such discussions, because of the potential benefits to having a field that is representative of and open to the diversity of our society. These issues are inherently cultural in nature; as such, how they are viewed be a function of the different disciplinary and national cultures engaged in an interdisciplinary and international endeavor, such as this workshop. That said, the effects of discrimination are severe and well-documented. A report recently outlined the barriers to access for women to permanent astronomy positions in France \citep{berne2020inequalities}. The American Astronomical Society Task Force on Diversity and Inclusion in Astronomy Graduate Education has published a report discussing strategies to improve the diversity and fairness in gradschool education \citep{AASDEI2018}. The US National Academy of Sciences, Engineering, and Medicine published a workshop report on the impacts of racism on Black people in sciences and engineering \citep{national2020impacts}, a report on the impacts of race and ethnicity on health care \citep{nelson2002unequal}, and of the prevalence and impacts of sexual harassment across academia \citep{national2018sexual}. They also provided a top-level strategy for ``reducing barriers to scientific excellence" in their Exoplanet Science Strategy. That report included the finding that ``Development and dissemination of concrete recommendations to improve equity and inclusion and combat discrimination and harassment would be valuable for building the creative, interdisciplinary teams needed to maximize progress in exoplanet science over the coming decades" \citep{WangDEI:2019}. If our field can make and follow such recommendations, it would likely generate improvements to our work, as suggested by other research; for example, increased diversity is shown to lead to an improvement in the productivity and outputs from groups and reorganizations \citep{page2008difference}, and cultures of inclusivity bring about an improvement of morale and a decrease in conflict \citep{nishii2013benefits}.

The need for inclusivity also extends to academic and disciplinary considerations. This research exists at the overlap between Earth sciences, astronomy, planetary sciences, and heliophysics. Incorporating the perspectives of these different disciplines is critical to success. Similarly, this research communities is global in extent, with teams conducting GCM simulations in many countries across multiple continents. Finally, this research faces any workforce challenges that other work in academic is presented with.

Based on this research, the workshop organizers believed that increasing the field in these ways will increase the variety of perspectives in our work, which will ultimately serve to improve the outputs from the community. They considered ways to ensure the workshop would do this along multiple axes of diversity, including but not limited to disciplines, institutions, genders, races, ethnicities, sexual orientations, disability statuses, cognitive diversity, nationalities, political affiliation, career stages, generations, job ranks, and levels of professional stability. Each of these aspects of diversity will require consideration on their own; in turn, action on any of them will also serve to lessen the negative impacts of discrimination in other areas. The workshop organizers included such considerations into the very structure of the meeting. To ensure accessibility across a global community, working in the context of a pandemic, they recorded lectures and made them available for later viewing and created a slack space for asynchronous communication. To account for the interdisciplinary scope of the meeting, the organizers posted introductory talks before the start of the meeting, to familiarize everyone with terminology and tools. To prevent harassment, participants agreed to adhere to a code of conduct that was shared on the meeting's home page. And the workshop included discussions of diversity in the field. One difficult issue was to determine how to structure these conversations. Originally, this discussion was scheduled as a ``breakout discussion" run in parallel with scientific/technical breakout discussions. However, some participants suggested we instead hold this as a ``whole group" discussion so that everyone would be engaged in the conversation, and so that those wanting to work on these efforts did not have to ``trade" discussions of these issues against technical/research discussions they also wanted to engage in. In response to this feedback, we dedicated time for the entire workshop community to discuss these issues, even though that came at the costs of a disruption to the planned schedule and less time for the breakout sessions.

Our discussions on diversity were organized around the idea of appreciative inquiry, where individuals share stories of past successes. In this case, we discussed ``a time when you were part of a diverse team in early career, which really benefited from its diversity." We asked about the environments in which that success was found, to highlight those instances of success. This discussion highlighted a number of areas of past success; we relate some of those examples here. In situations where there was good diversity along one or more axis, it helped value other areas of diversity as it also nurtures a sense of inclusivity in the organization. One participant claimed that leadership played a positive role in groups they had previously been a part of, and that good leadership helped the group advance their degree of inclusivity in that community. Other participants discussed the relevancy of current bridge initiatives that are already in place that have proven to be successful to help bring underrepresented minorities and low income students to the STEM workforce \citep{Crouse2020}. This further developed conversations about future bridge programs in the making that we expect to positively impact our community. The highly interdisciplinary nature of our THAI community could directly benefit from adopting similar pipelines and mentoring strategies of bridge programs and will ensure a more inclusive and representative community in the following years. Smaller group discussions were noted as helpful, as they gave voice to the perspectives of different backgrounds. In some cases, the diversity in a group provided extrinsic value, such as when a TA (teaching assistant) spoke the same languages as students and helped them learn class material, or differing perspectives produced better results. There was also discussion of open recruitment for positions, and for selection criteria centered on underlying skills, not statistics such as GPA or citations. General strategies that were discussed included training the next generation of role models, being an ally to people from underrepresented groups, and acknowledging both the real progress we have made and the challenges we have yet to address.

There is a related issue raised at our meeting that our community must also grapple with: that of equity for individuals without tenured positions. This group includes people in non-tenure-track faculty or research roles, as well as tenure-track faculty who have not yet received tenure. The community of non-tenured researchers is growing, both in real terms and as a percentage of our fields. As a result, the discrepancies in salary, financial security, and privileges in the workplace are increasing in their impacts on our ability to do this work \citep{Bourne8647,SSFNRIG2017}. The related stresses have impacts on the morale in our field, and it reduces the flexibility people have to spend time on these endeavors, which may not be the ones that lead to promotion. Additionally, at some institutions these discrepancies can block access to resources - such as funding for community service work. In that context, it can create the paradoxical situation where the people that have the time to conduct intermodel comparison simulations are the ones that do not have the funding or the professional stability for that activity. Specifically, workshop discussions highlighted that GCMs are very complex tools with generally steep learning curves for building, running, and modifying the codes.  They also  converge after days, weeks or sometimes months of computation, and can produce GB to TB in output to sort through.  While Earth climate science departments are familiar with these timescales and expectations, the intersection of climate modeling with the fast paced and hypercompetitive environment of exoplanet science and astronomy can prove challenging in terms of career advancement metrics.   In a very competitive field where the scientific productivity as an early career scientist is crucial, being a GCM modeler may be inhibiting, due to the long timescale to produce good and original science. This is especially true if they are similarly compared and evaluated to, for instance, observers that have a higher rate of publications and discoveries. Like other aspects of diversity and inclusion, this issue's impact can compound with other axes of power and privilege, and also leave individuals without the energy and career stability needed to address other aspects of diversity.   

These discussions were short, so the above approaches are a small subset of what is needed to improve the field. However, they provide a starting point for the necessary, sustained discussion on this topic. This will ultimately require thinking vertically across career stages, to develop a pipeline that allows people from any background the opportunity to join and meaningfully contribute to our field. We must then ensure those various backgrounds are included in our intellectual discussions and work, with intentional organization of open and inclusive conversations. We must work to ensure both formal and informal policies in the field are anti-discriminatory in nature. And our institutions need to do better to ensure equity and opportunity for people from all these backgrounds, and to people from different career stages and levels of job security.


\section{Conclusions of the Workshop and Perspectives}\label{sec:end}

The THAI workshop has allowed the exoplanet GCM community, focused on terrestrial planets, to discuss the role of GCMs in exoplanet characterization. THAI has been used as a vector in discussions between the various GCM groups (ExoCAM, LMD-G, ROCKE-3D, UM, THOR, Isca, etc.). From the THAI experiment, it is clear that clouds are the largest source of differences between the models. The average altitude of clouds and their optical thickness at the terminator affect the continuum level of the simulated transmission spectra. Various continuum levels therefore imply different detectability of molecular absorption lines, thereby impacting predictions of the detectability of an atmosphere with future space observatories such as JWST. Three papers are currently in preparation to present the THAI results and will be included within a focus issue ``Collection of model papers for GCM, EBM and 1D models applied to THAI" in the Planetary Science Journal (PSJ) alongside this workshop report.

The future of exoplanet GCMs will likely require the use of a hierarchical approach (i.e. simulations performed on a local grid in order to derive parameterizations of sub-grid processes to be used in low spatial resolution GCM simulations) and will not necessarily lean toward higher spatial resolutions. In addition, the workshop participants have identified clouds/hazes and convection as the first and second most important processes for the field to focus on in the upcoming years. 

GCMs do not have to be used alone - a scientific approach using a hierarchy of models such as EBMs, 1D radiative-convective models and GCMs is the key to progress efficiently on prediction observation and interpret data. However, GCM simulations are computationally expensive and - in a world where the climate is globally changing - the CO$_2$ emissions released by heavy computing should be controlled with strategies to reduce these emissions at a community level.

THAI has also demonstrated the utility of intermodel comparison for exoplanet science. To continue this initiative, we have proposed the Climates Using Interactive Suites of Intercomparisons Nested for Exoplanet Studies (CUISINES) that will host additional intercomparisons among exoplanet characterization studies in the future. A formal workshop on best practices for such intercomparisons will be organized in Fall 2021 to optimize the collaboration and science returns of CUISINES. 

If we wish to successfully grow our understanding of the Earth and the worlds beyond our own atmosphere, we need to ensure the GCM community reaches more diverse audiences. We hope that implementing Diversity \& Inclusion initiatives - such as bridge programs - will help move the scientific community forward in a way that brings equitable collaborations in the coming years.

\acknowledgments
The workshop SOC acknowledge funding support from the Nexus for Exoplanet System Science (NExSS). 
T. Fauchez, R. Kopparapu, S. Domagal-Goldman and M.J. Way acknowledge support from the GSFC Sellers Exoplanet Environments Collaboration (SEEC), which is funded in part by the NASA Planetary Science Divisions Internal Scientist Funding Model. M. Turbet received funding from the European Research Council (ERC) under the European Unions Horizon2020 research and innovation program (grant agreement No.  724427/FOUR ACES) and from the European Unions Horizon 2020 research and innovation program under the Marie Sklodowska-Curie Grant Agreement No.  832738/ESCAPE. M.T. thanks the Gruber Foundation for its generous support. This work has been carried out within the framework of the National Centre of Competence in Research PlanetS supported by the Swiss National Science Foundation. M.T. acknowledges the financial support of the SNSF.  M.J. Way and L. Sohl acknowledge funding support through NASA's Nexus for Exoplanet System Science (NExSS) via the ROCKE-3D working group. E.T. Wolf acknowledges funding support through NASA's Nexus for Exoplanet System Science (NExSS) via the Virtual Planetary Laboratory and the ROCKE-3D working groups.
G. Gilli acknowledges funding by the European Union's Horizon2020 research and innovation programme under the Marie Sklodowska-Curie Grant Agreement No. 796923/Hot-TEA. 
M. Lef\`evre acknowledges funding from the European Research Council (ERC) under the European Union’s Horizon 2020 research and innovation program (grant agreement No. 740963/EXOCONDENSE).
French co-authors were granted access to the High-Performance Computing (HPC) resources of Centre Informatique National de l'Enseignement Supérieur (CINES) under the allocations No. A0060110391 and A0080110391 made by Grand Équipement National de Calcul Intensif (GENCI). 

D.E.S., N.J.M. and I.A.B acknowledge use  of the Monsoon system, a collaborative facility supplied under the Joint Weather and Climate Research Programme, a strategic partnership between the Met Office and the Natural Environment Research Council. Some of this work was performed using the DiRAC Data Intensive service at Leicester, operated by the University of Leicester IT Services, which forms part of the STFC DiRAC HPC Facility (www.dirac.ac.uk). The equipment was funded by BEIS capital funding via STFC capital grants ST/K000373/1 and ST/R002363/1 and STFC DiRAC Operations grant ST/R001014/1. DiRAC is part of the National e-Infrastructure. This research also made use of the ISCA High Performance Computing Service at the University of Exeter. We acknowledge support of the Met Office Academic Partnership secondment programme.  This work was partly supported by a Science and Technology Facilities Council Consolidated Grant (ST/R000395/1).
We would like to thank Dorian S. Abbot and the anonymous reviewer for comments that greatly improved our manuscript.

\software{{\sc matplotlib} \citep{Hunter2007}}


\clearpage
\appendix

\section{Appendix information}
\subsection{Presentation of the THAI GCMs} \label{subsec:GCMs}
In this section we briefly review the four primary 3D climate models used in the THAI project: ExoCAM, LMD-G, ROCKE-3D, and UM.

\subsubsection{ExoCAM} \label{subsubsec:ExoCAM}

ExoCAM is an exoplanet branch of the Community Earth System Model (CESM) version 1.2.1.  CESM is provided publicly by the National Center for Atmospheric Research in Boulder, CO, (\url{http://www.cesm.ucar.edu/models/cesm1.2/})  and ExoCAM is freely available on GitHub (\url{https://github.com/storyofthewolf/ExoCAM}).  To use ExoCAM, the user must first obtain CESM v1.2.1, and then ExoCAM is installed as a patch on top of the core CESM code.  ExoCAM was developed by E.T. Wolf to facilitate accessible configurations for exoplanet and planetary modeling, and is now used by several different research groups in the community.  The ExoCAM code package includes model configurations, initial condition files, source code modifications, and an accompanying flexible correlated-k radiative transfer model, ExoRT  \url{https://github.com/storyofthewolf/ExoRT}).  ExoRT can be run coupled to the 3D model or in a standalone 1D mode and has several supported gas absorption schemes.   Typically, ExoCAM is run utilizing the cloud and convection physics from the Community Atmosphere Model (CAM) version 4 \citep{Neale2010}, and a finite volume dynamical core \citep{LinRood1996}; however, ExoCAM can be configured to leverage other CESM supported dynamical cores (e.g., spectral element cubed-sphere) and physics routines as desired (e.g., CAM5, CARMA).  Likewise, ExoCAM is most often run using a 4$^{\circ}$ $\times$ 5$^{\circ}$ horizontal resolution and 40 vertical atmospheric layers  up to 1 mbar pressures; however, ExoCAM can easily be run with other supported model resolutions (e.g., \citet{Wei2020}) and model tops (e.g., \citet{Suissa2020}) with relative ease.  For the THAI simulations, ExoCAM was run with 4$^{\circ}$ $\times$ 5$^{\circ}$ horizontal resolution, 51 vertical layers extending to 0.01 mbar pressures, configured with CAM4 cloud and convection physics, and with the ExoRT radiation scheme originally developed for Archean Earth atmospheres described in \citet{Wolf&Toon2013}. ExoCAM, coupled to ExoRT, has been used to study a variety of problems including deep paleoclimates for Earth \citep{Wolf&Toon2013, Wolf&Toon2014}, stellar and CO$_2$ driven moist greenhouse climates \citep{Wolf2015, Wolf2018}, the climate of Earth-like exoplanets around solar type stars \citep{Wolf2017, Adams2019b, Kang2019a, Kang2019b, Kang2019c}, tidally locked exoplanets around M-dwarf stars \citep{Kopparapu2017, Komacek&Abbot2019, YangH2019, Komacek2019, Komacek2020, Komacek2020b, Wei2020, Suissa2020, Zhang2020, Rushby:2020}, and Earth-like planets in circumbinary systems \citep{Wolf2020}.

\subsubsection{LMD-G} \label{subsubsec:LMD-G}

The LMD-G GCM - or the LMD Generic model - is a 3D Global Climate Model historically developed at the Laboratoire de Meteorologie Dynamique (LMD) in Paris, France. The model originally derives from the LMDz three-dimensional Earth \citep{Hourdin:2006} and Mars \citep{Forget:1999} Global Climate Models, but it benefits from the lessons learned by developing GCMs for most atmospheres in the solar system, where models can be tested against a wide range of observations \citep{Forget_lebonnois2013}. It solve the primitive equations of geophysical fluid dynamics using a finite difference dynamical core on an Arakawa C grid. The LMD-G GCM is equipped with flexible radiative transfer (based on the correlated-k method,  \citet{Word:10gj581}) and thermodynamics/cloud microphysics packages with the objective of being able to simulate any cocktail of atmospheric gases (as long as spectroscopic datasets are available) and aerosols. In particular it can account for the condensation of both minor and major constituents of an atmosphere. Most planetary (planet size, mass, rotation period, topography, etc.) and stellar parameters (insolation, input spectrum) can be easily adapted for a wide range of planets. LMD-G has been used in many climate studies for the past and future climates of solar system planets \citep{Forget:2013,Wordsworth:2013,Charnay:2013,Leconte:2013b,Charnay:2014,Turbet:2017epsl,Turbet:2017icarus,Turbet:2020} and exoplanets in a wide range of conditions \citep{Wordsworth:2011,Leconte:2013,Charnay:2015a,Turbet:2016,Bolmont2016,Charnay:2020}. It was specifically adapted and used for the TRAPPIST-1 planets in \citet{Turbet:2018aa} and \citet{Fauchez:2019}. More information on the model (code, user manual, tools, publications) can be found on \url{http://www-planets.lmd.jussieu.fr/}.
Recently, the flexible physical parameterizations of LMD-G have also been interfaced with LMD's next-generation icosahedral dynamical core DYNAMICO \citep{Dubos2014}, which is particularly suitable for massively-parallel architectures. DYNAMICO has been used to perform high-resolution simulations of solar system's giant planets \citep{Spiga2020,Caba:20,Bard:21} and carries many promising perspectives for exoplanet studies \citep{SainsburyMartinez2019}.

\subsubsection{ROCKE-3D} \label{subsubsec:ROCKE-3D}
The Resolving Orbital and Climate Keys of Earth and Extraterrestrial Environments with
Dynamics (ROCKE-3D) is a GCM developed at NASA Goddard Institute of Space Studies (GISS) \citep{Way2017}. ROCKE-3D is based on its parent Earth climate GCM GISS ModelE2 \citep{Schmidt2014} which is used for the Coupled Model Intercomparison Project
Phase 6 (CMIP), currently in its 6$\textsuperscript{th}$ version.
For the THAI intercomparison ROCKE-3D version Planet\_1.0 was used.
ROCKE-3D Planet\_1.0 was run at an atmospheric horizontal resolution of 4$^{\circ}$ $\times$ 5$^{\circ}$ with 40 vertical atmospheric layers. The atmospheric model top was 0.1 mb ($\sim$60 km altitude).  Since ROCKE-3D is an extension of its parent Earth model, it brings along many features of the parent model including river and underground runoff, ground hydrology for different soil types, and a dynamic lakes mode where lakes can either accumulate or dissipate depending upon the competition between evaporation and precipitation. The Planet\_1.0 version of ROCKE-3D used in the THAI intercomparison is extensively documented in \cite{Way2017} where one can find a more detailed description of its capabilities, features, and limitations. One important area where ROCKE-3D differs from its parent model comes from its use of a completely different radiative transfer scheme called SOCRATES (see Section \ref{subsubsec:UM}) which offers far more flexibility than the default GISS scheme. At the same time, SOCRATES is more computationally demanding than the default GISS scheme. Whereas the GISS scheme was designed to be extremely fast, it is only available for use with modern Earth atmospheric pressures and gas mixing ratios.
ROCKE-3D coupled to SOCRATES has been used in a variety of climate studies for solar system planets through time \citep{Way2016,DelGenio2018,DelGenio2020,Way2020} and beyond \citep{Way2017b,Way2018,Kane2018,Colose2019,DelGenio2019,Aleinov2019,Olson2020}.

\subsubsection{UM} \label{subsubsec:UM}
The Unified Model ({\sc UM}) has been developed by the UK Met Office and UM Partnership over the last 30 years with the aim of being able to use the same model for both operational weather forecasting and climate simulation. The {\sc UM} can be run with a range of planetary parameters, spatial and temporal resolutions, in global \citep{Walters2019} or regional \citep{Bush2020} configurations.
The {\sc UM}'s dynamical core solves the equations of motion using a semi-implicit, semi-Lagrangian method \citep{Wood2014}, with variables discretized on an Arakawa-C grid in the horizontal, and a staggered height-based terrain-following Charney-Phillips grid in the vertical.
The dynamical core is capable of solving a range of dynamical equations from the most simplified primitive equations, to those close to the full non-hydrostatic equations for a compressible fluid \citep[see][]{white_2005,Mayne_2014}.
The {\sc UM} includes sophisticated physical parameterizations for subgrid-scale turbulence, convection, water cloud and precipitation, as well as radiative transfer which is solved by the open-source, two-stream, correlated-$k$ code SOCRATES, accessible at \url{https://code.metoffice.gov.uk/trac/socrates} and surface and sub-surface processes\footnote{{\sc JULES} (\url{https://jules.jchmr.org/})}.
Note that through the SOCRATES radiative transfer code, the {\sc UM} is capable of generating synthetic spectra for any given 3D simulation \citep[e.g.,][]{Lines_2018b,Boutle_2020}.

Adaptation and application of the {\sc UM} to exoplanets has been led by the Exeter Exoplanet Theory Group (EETG, \url{exoclimatology.com}).
The {\sc UM} was initially benchmarked against a range of standard Earth-like planet tests \citep{Mayne_2014b}, opening an avenue for studying temperate extraterrestrial atmospheres.
Using the {\sc UM}, various climate processes in the atmospheres of tidally locked rocky exoplanets have been studied, focusing on the impacts of planet eccentricity and atmospheric composition \citep{Boutle2017}, size and location of a substellar continent \citep{Lewis_2018}, treatment of convection \citep{Sergeev2020}, host star spectrum \citep{Eager_2020}, presence of mineral dust \citep{Boutle_2020} and an interactive ozone cycle \citep{Yates_2020}.

The versatility of the {\sc UM} allowed for its application to a range of gas giant atmospheres, primarily $H$/$He$-dominated hot Jupiters.
After a successful adaptation the radiative transfer code \citep{Amundsen_2014,Amundsen_2017}, the {\sc UM} was used to explore flow structures in hot Jupiter atmospheres \citep{Mayne_17,Debras_2019,Debras_2020}, and to demonstrate that for smaller planets with extended atmospheres (i.e. mini-Neptunes, see Sec.~\ref{subsec:limdyn}) the often used primitive equations may not accurately capture the atmospheric dynamics \citep{Mayne_2019}.
Additionally, a flexible gas-phase chemistry scheme \citep{Drummond_2016} was coupled to the {\sc UM} allowing for 3D simulations of $H$/$He$-dominated atmospheres with both equilibrium \citep{Drummond_2018} and kinetic \citep{Drummond_2020} gas--phase chemistry.
To parameterize clouds in the atmospheres of hot Jupiters, the {\sc UM} was also coupled to both a detailed high-temperature microphysics scheme \citep{Lines_2018b,Lines_2018} and a steady-state simplified cloud scheme \citep{Lines_2019}.

\bibliography{report.bib}
\bibliographystyle{aasjournal}



\end{document}